\documentclass[11pt]{article}
\usepackage{epsfig}
\usepackage{amsfonts}
\usepackage{amsmath}
\usepackage{amssymb}
\usepackage{bbm,bm}
\usepackage{cite}
\allowdisplaybreaks[1]

\usepackage{tikz}
\usetikzlibrary{shapes,arrows,shadows,automata,positioning}
\newdimen\nodeDist
\usepackage{comment}
\usepackage[normalem]{ulem} 

\usepackage[
	colorlinks=true,
	linkcolor=black,
	citecolor=black,
	filecolor=black,
	urlcolor=black,
        breaklinks=true
        ]{hyperref}
\usepackage{multicol}

\renewcommand{\vec}[1]{{\bf #1}}

\newcommand{\Lamd}{\Lambda_{\rmii{3d}}}

\newcommand\MSbar{$\overline{\rm MS}$}

\newcommand{\gY}{g_\rmii{$Y$}}

\newcommand{\gp}{g'}

\newcommand{\sumint}[1]{\hbox{$\sum$}\!\!\!\!\!\!\!\int_{#1}}

\def\lsi{\raise0.3ex\hbox{$<$\kern-0.75em\raise-1.1ex\hbox{$\sim$}}}
\def\gsi{\raise0.3ex\hbox{$>$\kern-0.75em\raise-1.1ex\hbox{$\sim$}}}

\newcommand{\nn}{\nonumber \\}
\newcommand{\rmi}[1]{{\mbox{\scriptsize #1}}}
\newcommand{\rmii}[1]{{\mbox{\tiny\rm{#1}}}}

\newcommand{\Tint}[1]{{\hbox{$\sum$}\!\!\!\!\!\!\!\int\,}_{\!\!\!\!\raise-0.9ex\hbox{$\scriptstyle{#1}$}}}
\newcommand{\Tinti}[1]{{{\Sigma}\!\!\!\!\raise0.3ex\hbox{$\int$}_\rmii{${#1}$}}}
\newcommand{\Tintip}[1]{{{\Sigma'}\!\!\!\!\!\raise0.3ex\hbox{$\int$}_\rmii{${#1}$}}}

\newcommand{\deltabar}{\raise-0.02em\hbox{$\bar{}$}\hspace*{-0.8mm}{\delta}}

\newcommand\ct{c_{\bar{\theta}}}
\newcommand\st{s_{\bar{\theta}}}

\makeatletter \@addtoreset{equation}{section} \makeatother
\renewcommand{\theequation}{\arabic{section}.\arabic{equation}}
\makeatletter
\renewcommand\section{\@startsection{section}{1}{\z@}%
  {-5.5ex \@plus -1ex \@minus -.2ex}
  {2.3ex \@plus.2ex}%
  {\normalfont\large\bfseries}}
\renewcommand\subsection{\@startsection{subsection}{2}{\z@}%
  {-3.25ex\@plus -1ex \@minus -.2ex}%
  {1.5ex \@plus .2ex}%
  {\normalfont\normalsize\bfseries}}
\renewcommand\thesection{\@arabic\c@section}
\renewcommand\thesubsection{\thesection.\@arabic\c@subsection}
\renewcommand{\@seccntformat}[1]{%
  \csname the#1\endcsname.\hspace{1.0em}}
\makeatother

\begin{document}

\flushbottom

\begin{titlepage}

\begin{flushright}
NORDITA 2022-031\\
DESY-22-091\\
\end{flushright}
\begin{centering}

\vfill

{\Large{\bf
Speed of sound in cosmological phase transitions and \\ effect on gravitational waves
}}

\vspace{0.8cm}

\renewcommand{\thefootnote}{\fnsymbol{footnote}}
Tuomas V.~I.~Tenkanen$^{\rm a,b,c}$%
\footnote{tuomas.tenkanen@su.se} and
Jorinde van de Vis$^{\rm d,}$%
\footnote{jorinde.van.de.vis@desy.de}

\vspace{0.8cm}

$^\rmi{a}$%
{\em
Nordita,
KTH Royal Institute of Technology and Stockholm University,\\
Roslagstullsbacken 23,
SE-106 91 Stockholm,
Sweden\\}
\vspace{0.3cm}

$^\rmi{b}$%
{\em
Tsung-Dao Lee Institute \& School of Physics and Astronomy, Shanghai Jiao Tong University, Shanghai 200240, China\\}

$^\rmi{c}$%
{\em
Shanghai Key Laboratory for Particle Physics and Cosmology, Key Laboratory for Particle Astrophysics and Cosmology (MOE), Shanghai Jiao Tong University, Shanghai 200240, China\\}

$^\rmi{d}$%
{\em
Deutsches Elektronen-Synchrotron DESY, Notkestr. 85, 22607 Hamburg, Germany\\}

\vspace*{0.8cm}

\mbox{\bf Abstract}

\end{centering}

\vspace*{0.3cm}

\noindent

The energy budget for gravitational waves of a cosmological first order phase transitions depends on the speed of sound in the thermal plasma in both phases around the bubble wall. Working in the real-singlet augmented Standard Model, which admits a strong two-step electroweak phase transition, we compute higher order corrections to the pressure and sound speed. We compare our result to lower-order approximations to the sound speed and the energy budget and investigate the impact on the gravitational wave signal. We find that deviations in the speed of sound from $c_s^2 = 1/3$ are enhanced up to $\mathcal O(5\%)$ in our higher-order computation. This results in a suppression in the energy budget of up to $\mathcal O (50\%)$ compared to approximations assuming $c_s^2 = 1/3$. The effect is most significant for hybrid and detonation solutions. We generalise our discussion to the case of multiple inert scalars and the case of a reduced number of fermion families in order to mimic hypothetical dark sector phase transitions. In this sector with modified field content, the sound speed can receive significant suppression, with potential order-of-magnitude impact on the gravitational wave amplitude.

\vfill
\end{titlepage}

\tableofcontents
\clearpage

\renewcommand{\thefootnote}{\arabic{footnote}}
\setcounter{footnote}{0}

%
\section{Introduction}
\label{sec:intro}

\paragraph*{Gravitational waves from first order phase transitions}
A stochastic gravitational wave (GW) background generated by a cosmological phase transition can offer a window to the early universe predating the recombination era. 
In a first order phase transition, a barrier separates the two degenerate
minima of the free energy of the thermal plasma at the critical temperature $T_c$. When the temperature ($T$) of the plasma drops below the critical temperature, the system transitions from the high temperature to the lower temperature phase via thermal or quantum fluctuations. Bubbles of the low-$T$ phase nucleate, expand due to the pressure difference between the phases inside and outside the bubble, and collide eventually, filling the whole universe with the low-$T$ phase. A fraction of the latent heat of the transition gets converted into kinetic energy in the form of sound waves within the plasma, which act as a source of shear stress producing gravitational waves, resulting in a stochastic background \cite{Hogan:1986qda,Hindmarsh:2013xza,Caprini:2015zlo,Hindmarsh:2015qta,Hindmarsh:2017gnf,Caprini:2019egz}.
A detection of such a hypothetical background is one of the science goals of the space-based GW observatory LISA \cite{2017arXiv170200786A} and other similar next-generation GW experiments.  

The Standard Model (SM) of particle physics does not predict any cosmological first order phase transitions, as both the electroweak (EW) and QCD phase transitions occur via a smooth crossover \cite{Kajantie:1996mn,Kajantie:1995kf, Csikor:1998eu, Stephanov:2006zvm}.
Hence, the existence of a cosmic GW background would point to physics beyond the Standard Model (BSM). The GW signal would offer a probe complementary to high energy colliders \cite{Alves:2018jsw,Chala:2018opy,Ramsey-Musolf:2019lsf}.
Furthermore, a first order EW phase transition is a crucial ingredient in EW baryogenesis, a dynamical mechanism 
aiming to explain the matter/antimatter imbalance of the present day universe \cite{Kuzmin:1985mm,Morrissey:2012db,Konstandin:2013caa,White:2016nbo,Bodeker:2020ghk}.
Multiple BSM theories have been proposed that are believed to incorporate cosmic phase transitions, 
see e.g. the references listed 
in \cite{Ramsey-Musolf:2019lsf,Schicho:2021gca}. 

\paragraph*{Modelling the signal -- importance of the sound speed}
The gravitational wave signal of a cosmological phase transition receives contributions from the kinetic energy stored in the bubble walls \cite{Jinno:2017fby, Konstandin:2017sat, Cutting:2018tjt, Cutting:2020nla}, sound waves in the plasma, and turbulence \cite{Caprini:2009yp, Niksa:2018ofa, RoperPol:2019wvy, Kahniashvili:2020jgm, Auclair:2022jod}.
In the case of weak to moderate phase transitions, the sound wave contribution is expected to be dominant \cite{Hindmarsh:2013xza,Caprini:2015zlo,Caprini:2019egz},
and it will be the main focus of this work.

The gravitational wave signal from sound waves has been determined in a range of numerical simulations \cite{Hindmarsh:2013xza,Hindmarsh:2015qta,Hindmarsh:2017gnf, Cutting:2019zws}. The results of \cite{Hindmarsh:2015qta,Hindmarsh:2017gnf} were used to obtain a fitting function \cite{Caprini:2019egz} which approximates the gravitational wave signal as a function of the nucleation temperature ($T_n$), the inverse transition duration ($\beta$), the bubble wall speed ($v_w$), the sound speed ($c_s$) -- which is obtained from the pressure, and usually assumed to be $c_s \sim 1/\sqrt 3$ -- and the kinetic energy fraction, or energy budget, $K$. The value of $K$ can be determined from the fluid velocity profile of an isolated bubble, but is usually obtained from a fit provided by \cite{Espinosa:2010hh}, which gives the value of $K$ as a function of the phase transition strength $\alpha$ and $v_w$. 

The results of \cite{Espinosa:2010hh} were obtained in the `bag model', which assumes that both the broken and the symmetric phase are composed of radiation degrees of freedom only. The same assumption underlies the choice of $c_s = 1/\sqrt 3$. In a more realistic model, these assumptions do not hold, as some degrees of freedom are usually massive around the phase transition temperature.
The procedure sketched above thus neglects some of the model-dependence of the signal, by describing it fully in terms of $T_n, \alpha, \beta$ and $v_w$. This leads to an inaccuracy in the estimated gravitational wave signal, but also
possibly to an unnecessary degeneracy between different models.

Recently, it was demonstrated in \cite{Giese:2020rtr,Giese:2020znk} that further model-dependence of $K$ can, to a good approximation, be captured by a modification of the phase transition strength (see Eq.\eqref{eq:pseudotrace} below) and the values of the speed of sound of both phases of the phase transition. A code snippet provided by \cite{Giese:2020znk} serves as an alternative to the fit based on the bag model by \cite{Espinosa:2010hh}. It was demonstrated for a set of toy models, that the code snippet approximates the numerically obtained energy budget with an error of $ \lesssim 5\%$, compared to an error of $ \lesssim 50\%$ for a mapping onto the bag model.

Although some studies of the effect of the sound speed on the shape and strength of the gravitational wave signal now exist \cite{Giese:2020rtr,Giese:2020znk,Wang:2020nzm,Wang:2021dwl}, an accurate computation of the \emph{value} of the speed of sound in a first order phase transition is still lacking. Careful studies of the sound speed thus far have been limited to QCD \cite{Laine:2006cp,HotQCD:2014kol} and the SM electroweak crossover \cite{Laine:2015kra}.  
The goal of this work is to improve the accuracy of the computation of the pressure 
-- for a representative BSM model that accommodates a strong electroweak or dark sector phase transition \cite{Croon:2018erz} --
and in particularly to determine what are realistic values of the sound speed. To understand the importance of our analysis, we will compare the resulting amplitude of the gravitational wave spectrum with the result obtained in the approximation $c_s = 1/\sqrt 3$. 
In particular, we will study the SM extended by $N$ singlets with $O(N)$ symmetry as a proxy for first order electroweak phase transitions with an extended Higgs sector. We will also investigate the effect of the removal of some of the light SM fermions, mimicking a dark sector with a smaller number of degrees of freedom. 

The reason that we include multiple inert singlets is that
one can expect deviations in the speed of sound when additional massive particles are present.
In particular we expect that the deviation from $c_s^2 = 1/3$ increases with increasing $N$. 
In addition, we also manually remove some of the light degrees of freedom by reducing number of fermion families ($N_f$) to one instead of three. In this case, one can expect the sound speed to drop, because the many light degrees of freedom all push the speed of sound (squared) towards 1/3. We indeed observe such behaviour, and can conclude that deviations in the speed of sound can be relevant for dark sector phase transitions.

\paragraph*{BSM model}
While the features that we discuss are generic for thermal phase transitions in BSM scalar extensions, practically we work in the model where the SM is augmented with an $O(N)$ symmetric scalar singlet \cite{Drozd:2011aa}.
The Langrangian reads
\begin{align}
\mathcal{L} &= \mathcal{L}_{\text{SM}} + \frac{1}{2} \mu^2_S S^{\text{T}} S + \frac{1}{4} \lambda_S (S^{\text{T}} S)^2 + \frac{1}{2} \lambda_m S^{\text{T}} S H^\dagger H, 
\end{align}
where the singlet decomposes as $S = (S_1, ... , S_N)^{\text{T}}$ such that each component $S_i$ is real. We will focus on a two-step phase transition.
We assume that in the first step of the phase transition only one of the components of the singlet
gets a non-zero vev and participates dynamically in the phase transition, while the other components remain inert. Therefore, for $N=1$ this setup matches the $Z_2$-symmetric xSM studied in 
\cite{Brauner:2016fla, Gould:2021dzl, Schicho:2021gca, Niemi:2021qvp} in the context of dimensionally reduced effective field theories. 
We generalise the results of these works by including the inert singlet contributions 
for general $N$. 
As another novelty, we also compute the coefficient of the unit operator that describes the hard and soft scale contributions to the symmetric phase pressure. These contributions were not discussed in earlier singlet dimensional reduction literature as therein the intrest was purely in the pressure difference between different phases, and corrections to the speed of sound were not determined.

\paragraph*{Outline of this work}
This article is organised as follows. In Sec.~\ref{sec:gwtheory} we review the setup for gravitational wave production, paying attention to the role of the sound speed. In Sec.~\ref{sec:pressure} we discuss the computation of the pressure and inclusion of several higher order corrections. In Sec.~\ref{sec:GWs} we present our results for quantities important for the GW wave power spectrum. In Sec.~\ref{sec:dark-sector} we discuss how our findings could generalise to phase transitions in a dark sector. In Sec.~\ref{sec:discussion} we discuss our conclusions and their impact. Appendix \ref{sec:details} collects multiple technical details of our analysis.

%
\section{Model dependence of the gravitational wave signal: pressure and speed of sound}
\label{sec:gwtheory}

For a BSM model with a first order phase transition, the gravitational wave signal from sound waves can be estimated as
\cite{Caprini:2019egz}
\begin{align}
	\frac{d \Omega_{{\rm gw},0}}{d \ln{(f)}} = 0.687 F_{{\rm gw},0} K^2 \left(H_* \tau_s \right) H_* R_*/c_s \tilde\Omega_{\rm gw} C\left(\frac{f}{f_{{\rm p},0}}\right) ,\label{eq:GWfit}
\end{align}
where $F_{{\rm gw},0}$ accounts for the redshift
from the generation of the signal until now, $H_*$ is the Hubble parameter at the phase transition temperature, 
 $R_*$ is the mean bubble separation
 at the moment of collision, given by $R_* \sim (8\pi)^{1/3} v_w/\beta$.
 $c_s$ is the speed of sound in the plasma, $\tilde \Omega_{\rm gw}$ is a numerical factor, $C$ is a function that determines the shape of the spectrum and $f_{{\rm p},0}$ is the peak frequency. 
 $\tau_s$, the lifetime of the source, is estimated as \cite{Ellis:2020awk}
\begin{align}
	\tau_s = 
	\begin{cases}
		1/H_*& \qquad H_* \tau_{\rm sh} < 1\, , \\
		R_*/K^{1/2}& \qquad H_* \tau_{\rm sh} > 1\, ,
	\end{cases}
\end{align}
where $\tau_{\rm sh}$ is the time of shock formation.\footnote{The shock formation time can be estimated as \cite{Pen:2015qta, Hindmarsh:2017gnf} \begin{equation*}
	\tau_{\rm sh} \sim \frac{R_*}{\bar U_f}\, ,\qquad \bar U_f = \left(\frac{K}{\Gamma} \right)^{1/2}\, ,
\end{equation*} where $\bar U_f$ is the enthalpy-weighted root-mean-square fluid velocity and $\Gamma \sim 4/3$ is the adiabatic index. 
} When shock formation is relatively slow, the gravitational wave spectrum scales with $K^2$, and for fast-developing shocks the spectrum scales as $K^{3/2}$. The kinetic energy budget $K$ is the fraction of energy that is available for the production of gravitational waves. It can be determined by solving the hydrodynamic equations for the plasma around a single expanding bubble. We will now give a summary of this computation, for a more complete description, see e.g. \cite{LandauLifshitz, Kamionkowski:1993fg, KurkiSuonio:1995pp, Espinosa:2010hh}.

The hydrodynamics follow from the energy-momentum tensor of the plasma:
\begin{align}
	T_{\mu\nu} = w u_\mu u_\nu - g_{\mu\nu} p,
\end{align}
where $u_\mu$ is the four-velocity of the plasma and $p$ and $w$ are the pressure and enthalpy respectively. The pressure is equal to the negative of the free energy of the plasma, and the enthalpy and energy density $e$ can be obtained from the pressure via
\begin{align}
	w = T\frac{\partial p}{\partial T}, \qquad e = T \frac{\partial p}{\partial T} -p.
\end{align}
The fluid equations are simply the continuity equations $\partial^\mu T_{\mu\nu}$. By projecting these along the directions parallel and perpendicular to the fluid flow, and by introducing the radial coordinate $\xi = r/t$, with $r$ the distance from the center of the bubble and $t$ the time since nucleation, the following hydrodynamic equations are found
\begin{align}
	\frac{d v}{d\xi} &= \frac{2v(1-v^2)}{\xi(1-v\xi)}\left(\frac{\mu(\xi,v)^2}{c_s^2} -1 \right)^{-1}, \label{eq:fluid1}\\
	\frac{dw}{d\xi} &= w \left(1 + \frac{1}{c_s^2} \gamma^2 \mu(\xi,v) \frac{dv}{d\xi} \right), \label{eq:fluid2}
\end{align}
where $v(\xi)$ is the fluid velocity, $\gamma$ the Lorentz factor and $\mu(\xi,v)$ the boosted velocity
\begin{align}
	\mu(\xi,v) = \frac{\xi - v}{1 - \xi v}.
\end{align}
The speed of sound can be derived from the pressure via
\begin{align}
	c_s^2 = \frac{dp/dT}{de/dT}.\label{eq:speedofsound}
\end{align}
We compute the sound speed separately for each metastable phase.%
\footnote{
Consequently,
 $c_s^2$ does not vanish at the transition point, in contrary to the prescription in \cite{Schmid:1998mx}.
}
The boundary conditions at the bubble wall are given by 
\begin{align}
	\frac{v_+}{v_-} & = \frac{e_b(T_-)+p_s(T_+)}{e_s(T_+) + p_b(T_-)}, \label{eq:bound1}\\
	v_+v_- & = \frac{p_s(T_+)-p_b(T_-)}{e_s(T_+) - e_b(T_-)} \label{eq:bound2},
\end{align}
where the $+(-)$ sign denotes a quantity right in front of (behind) the bubble wall. The subscript $s (b)$ refers to the equation of state of the symmetric (broken) phase. 
The boundary condition for $w$ is that $w = w_n$ in the region in front of the bubble where the fluid is at rest.
We use the subscript $n$ to denote quantities evaluated at the nucleation temperature $T_n$.
The kinetic energy fraction $K$ is obtained from the solution of the hydrodynamic equations via
\begin{align}
	K = \frac{3}{e_n v_w^3} \int d\xi \xi^2 v^2 \gamma^2 w, \label{eq:defK}
\end{align}
where $v_w$ is the bubble wall velocity. 

Although both the fluid equations~\eqref{eq:fluid1} and \eqref{eq:fluid2} and the boundary conditions \eqref{eq:bound1} and \eqref{eq:bound2} depend on the equation of state, Refs.~\cite{Giese:2020rtr,Giese:2020znk} show that this dependence can, to a good approximation, be captured by the following parameters only: the speed of sound in the broken phase $c_{s,b}$, the speed of sound in the symmetric phase $c_{s,s}$ (both evaluated at the nucleation temperature) and the phase transition strength
\begin{align} 
	\alpha_{\bar \theta} =\frac{D \bar \theta (T_n)}{3 w_n}, \qquad {\rm with} \quad \bar \theta = e - \frac{p}{c_{s,b}^2}, \qquad  {\rm and} \quad D\theta(T_n) = \theta_s(T_n) - \theta_b (T_n).\label{eq:pseudotrace}
\end{align}
Consequently, instead of solving the full hydrodynamic equations for each BSM model, $K$ can be obtained as a function of $c_{s,b}, c_{s,s}$, $v_w$ and $\alpha_{\bar \theta}$ in a simplified model using the code snippet provided by \cite{Giese:2020znk}. This procedure is a generalisation of the fit in terms of the phase transition strength and $v_w$ provided by Ref.~\cite{Espinosa:2010hh}, using the bag equation of state. The new approximation works particularly well when the temperature-dependence of the sound speed is weak, but approximates the full numerical solution to
an accuracy of
 $< 5 \%$ for all toy models considered in \cite{Giese:2020rtr,Giese:2020znk}. 

To quantify the importance of an accurate computation of the sound speed,
in Sec.~\ref{sec:GWs} we will determine the kinetic energy fraction for our benchmark point using three different methods (these methods are identical to methods 1-3 in \cite{Giese:2020znk}):
\begin{itemize}
	\item{Method 1: full solution of Eqs.~\eqref{eq:fluid1} and \eqref{eq:fluid2}, using the complete, temperature-dependent speed of sound.}
	\item{Method 2: solution of Eqs.~\eqref{eq:fluid1} and \eqref{eq:fluid2}, assuming constant sound speeds in the broken and symmetric phases.}
	\item{Method 3: assuming the speed of sound is $c_s = 1/\sqrt 3$. This corresponds to a mapping onto the bag equation of state. Note that this is the most common method in the literature, and it was also used in Ref.\cite{Croon:2020cgk}, where the uncertainty in the GW spectrum associated to the other thermodynamic quantities was estimated.}
\end{itemize}
The difference between Method 1 and Method 3 gives a measure of the importance of a careful computation of the speed of sound. The difference between Method 1 and Method 2 gives a measure of the temperature-dependence of the speed of sound.

In this work we will assume that the gravitational wave spectrum is fully described by the broken power law of Eq.~(\ref{eq:GWfit}), and only investigate the effect of our improved sound speed computation on the overall amplitude. According to Eq.~(\ref{eq:GWfit}), the position of the peak is determined by the size of the bubbles at the moment of collision. However, results of the sound shell model \cite{Hindmarsh:2016lnk, Hindmarsh:2019phv} and hybrid simulations \cite{Jinno:2020eqg} suggest that the width of the sound shells also gets imprinted onto the spectrum, resulting in a doubly broken power law shape. The results of \cite{Hindmarsh:2016lnk, Hindmarsh:2019phv,Jinno:2020eqg} were all obtained under the assumption that the sound speed is $c_s = 1/\sqrt 3$. Since deviations on the sound speed affect the shape of the fluid profile \cite{Leitao:2014pda, Giese:2020znk} it is expected that the shape of the gravitational wave spectrum is also affected by variations in the sound speed. This effect was demonstrated for the sound shell model in \cite{Wang:2021dwl}, but will not be considered further in our analysis.


\section{Computation of the pressure and sound speed}
\label{sec:pressure}

In this section we review how to compute the pressure perturbatively in high temperature field theory. We present this discussion for the SM augmented with $N$ singlets with O($N$) symmetry, but this discussion readily generalises to other BSM theories.
In fact, many of the computations presented in this work can be obtained with the automated {\tt DRalgo} package for generic, user defined models \cite{Ekstedt:2022bff}. 
Many details of our computation in the model in question are relegated to Appendix \ref{sec:details}.

\paragraph*{Using dimensional reduction to determine the phase transition parameters}
In order to improve the accuracy of the computation of the sound speed, we will use the method of high temperature dimensional reduction \cite{Ginsparg:1980ef, Appelquist:1981vg}. 
Therein, the perturbative computation is organised in the effective field theory language \cite{Kajantie:1995dw,Braaten:1995cm}, based on a chain of scale hierarchies at high temperature. In the rest of this article, we will refer to the different scales as hard ($\pi T$), soft ($g T$) and ultrasoft ($g^2 T$), where $g$ is a weak coupling, often identified with the weak gauge coupling in EW theories.%
\footnote{
In QCD the soft and ultrasoft scales are often referred to as the electric and magnetic scales, respectively.
}  
This method was recently applied in  \cite{Croon:2020cgk,Gould:2021oba} to obtain higher order corrections of $T_n$, $\alpha$ and $\beta$.  Ref.~\cite{Croon:2020cgk} reported an alarming leftover renormalisation scale dependence plaguing the conventional analysis based on a mere one-loop determination of the effective potential -- that describes the free energy and pressure of the plasma -- which in general signals slow convergence of perturbation theory.
In particular, there could be a multiple order-of-magnitude uncertainty in the predicted amplitude of the GW power spectrum, location of the peak frequency, and the consequent signal-to-noise (SNR) ratio for experiments such as LISA.
In \cite{Gould:2021oba}, it was reviewed and discussed in detail why conventional one-loop analyses fail to provide better accuracy: due to slower convergence of the perturbative expansion at high temperatures, several two-loop contributions have to be included to enable leading renormalisation group (RG) improvement.
After such RG improvement, order-by-order in perturbation theory, the renormalisation scale-dependence cancels between the running of lower order contributions and explicit logarithms at higher orders.
The inclusion of these two-loop level contributions is standard in a three-dimensional effective field theory (3d EFT) description of the phase transition thermodynamics, from which equilibrium properties of the transition, such as the pressure, and subsequent quantities of interest such as the critical temperature, latent heat, and the speed of sound can be derived. 

In this work at hand, we focus on the computation of the pressure and the speed of sound, and investigate the importance of its higher order corrections to the energy budget $K$, and hence to the GW power spectrum. In practice, we compute the pressure by following \cite{Gynther:2005dj,Gynther:2005av,Laine:2015kra}, and extend the SM computation of these works to a BSM setup. For a similar, related discussion of QCD and the speed of sound in a quark-gluon plasma, see e.g \cite{Laine:2006cp,HotQCD:2014kol} (c.f. also the review \cite{Ghiglieri:2020dpq}).
   
Although a necessary improvement in the accuracy of equilibrium quantities can be made by including higher order corrections to the pressure, or the effective potential, it was concluded in \cite{Gould:2021oba} that the most limiting source of theoretical uncertainty in the phase transition parameters is the bubble nucleation rate. This rate is used to derive $T_n$ and all other parameters have to be evaluated at this temperature. In \cite{Gould:2021oba}, the bubble nucleation rate was determined only at leading order within the 3d EFT. However, recent work has described how to extend such a computation to higher orders \cite{Ekstedt:2021kyx,Gould:2021ccf,Ekstedt:2022tqk,Ekstedt:2022ceo} (also c.f. \cite{Lofgren:2021ogg, Hirvonen:2021zej,Hirvonen:2022jba}). In the work at hand, we will not focus on such an improvement for the bubble nucleation rate, but instead treat $T_n$ as an input parameter that we vary; once its computation at higher orders is available, our results for the pressure, sound speed and the kinetic energy fraction can be used to determine the actual effect on GW power spectra.

\paragraph*{Pressure in perturbation theory}
In perturbation theory,
the pressure admits the following form and subsequent formal expansion%
\footnote{
The analogous expansion in QCD has been worked out, order by order:  $g^2$ \cite{Shuryak:1977ut, Chin:1978gj}, $g^3$ \cite{Kapusta:1979fh}, $g^4 \ln g$ \cite{Toimela:1982hv}, $g^4$ \cite{Arnold:1994ps, Arnold:1994eb}, $g^5$ \cite{Braaten:1995jr, Zhai:1995ac} and $g^ 6 \ln g$ \cite{Kajantie:2002wa,Kajantie:2003ax}.
Here, logarithmic contributions are highlighted separately, since they appear in a result expanded in $g$, when the 3d EFT formalism is not used. 
For the case of non-zero quark chemical potentials, see \cite{Vuorinen:2003fs}. 
} 
in the high temperature limit: 
\begin{align}
\label{eq:pressure-formal}
p(T) = p_{\text{sym}} - V_{\text{eff}}(v,s) \simeq T^4 \Big( a + b_1 g^2 + c g^3 + d g^4 + e g^5 + \mathcal{O}(g^6) \Big) + m^2 T^2 \Big( b_2 g^2 \Big),
\end{align}
where $p_{\text{sym}}$ describes the pressure in the symmetric phase where all scalar background fields vanish and $V_{\text{eff}}$ is the effective potential as a function of the background fields, and $a$-$f$ represent constants. 
The \MSbar{} mass parameters are assumed to 
scale as $m^2 \sim (g T)^2$.

We comment below on our conventions for the normalisations of both terms. A formal expansion in $g$
follows when a power counting rule is associated to each coupling. 
In particular we assume that all quartic couplings $\lambda$ scale as $g^2$, and we assume that there is a barrier present in the leading order effective potential; for radiatively generated transitions, the corresponding expansion is different \cite{Ekstedt:2022zro}.%
\footnote{
For example, for a radiatively generated transition, the 3d scalar mass parameter scales as $\mu^2_3 \sim g^3 T^2$, which leads to fractional powers $g^{4.5}$ and $g^{5.5}$ coming from the one-loop ultrasoft contribution \cite{Gynther:2005av}. 
}
Compared to the perturbative expansion of the effective potential at zero temperature, the structure of the perturbative expansion of the pressure is inflicted by several finite-$T$ peculiarities: odd powers of $g$ 
appear,
logarithmic terms include ratios of various thermal mass scales, and crucially the $\mathcal{O}(g^6)$ term is not even attainable in perturbation theory, due to the Linde problem \cite{Linde:1980ts}. 
In addition, the coupling expansion
misaligns with the loop expansion due to the enhancement of IR 
contributions. 

We emphasise, however, that in practice the pressure is not 
computed 
directly order-by-order in the above expansion in $g$. 
The reason is, that the pressure is not directly computable in such an expansion, since the direct use of perturbation theory is inhibited by various IR singularities. These have to be tackled by thermal resummations, based on a scale hierarchy of the different mass scales (hard/soft/ultrasoft) at high temperature. 
The perturbative result will include logarithms of all three mass scales in the problem, and it is not possible to choose a UV cutoff that would simultaneously remove all large logarithms.
In particular, these logarithms are of the form $\ln(\mu/T)$, $\ln(T/m_\rmii{E})$ and $\ln(m_\rmii{E}/m_\rmii{M})$, where $m_\rmii{E} \sim g T$ and $m_\rmii{M} \sim g^2 T$ are the soft and ultrasoft -- or electric (E)
 and magnetic (M) -- mass scales. These logarithms lead formally to orders $g^4 \ln(g)$ and $g^6 \ln(g^6)$, where logarithms are large for $g \rightarrow 0$.  

Dimensional reduction to a chain of EFTs resolves this problem systematically:
the thermal contributions of the hard scale are resummed as contributions to the parameters of the soft and ultrasoft scales.
The running within the EFT introduces new RG scales ($\mu_3$ and $\bar{\mu}_3$), which can be chosen independently, allowing to avoid large logarithms.
In our computation, we associate $p_{\text{sym}}$ with the {\it coefficient of the unit operator} \cite{Braaten:1995cm} in dimensional reduction, and this contribution describes the hard and soft contributions to the symmetric phase pressure. Contributions of the ultrasoft scale are, by our convention, encoded in $V_{\text{eff}}$, for both symmetric and broken phases.

Let us take a closer look at how different order terms in the pressure arise, and how we label them:
\begin{itemize}
\item[] LO ($a g^0$): one-loop hard contributions to $p_{\text{sym}}$. 
\item[] NLO ($b g^2$): tree-level terms in $V_{\text{eff}}$ and two-loop hard pieces in $p_{\text{sym}}$ . 
\item[] NNLO ($c g^3$): one-loop soft (ultrasoft) terms in $p_{\text{sym}}$ ($V_{\text{eff}}$). 
\item[] $\text{N}^3$LO ($d g^4$): two-loop (three-loop) soft (hard) pieces in $p_{\text{sym}}$  and two-loop ultrasoft pieces in $V_{\text{eff}}$. Tree-level $V_{\text{eff}}$ includes contributions at this order,
via the resummed parameters in the EFT.
\item[] $\text{N}^4$LO ($e g^5$): three-loop soft (ultrasoft) contributions to $p_{\text{sym}}$ ($V_{\text{eff}}$).
\end{itemize}
Note that terms that are formally of higher order are resummed inside the 3d EFT parameters in different contributions
and, in accord with the EFT construction, 
we will keep them untruncated in the final result \cite{Blaizot:2003iq,Laine:2006cp}.
We remind that the $\mathcal{O}(g^6)$-contribution is non-perturbative,
although the $\mathcal{O}(g^6 \ln g)$ logarithmic term is still computable at 4-loop level \cite{Kajantie:2002wa,Kajantie:2003ax}. 

The LO result for the sound speed, $c^2_s = 1/3$, follows from the LO result for the pressure. Typical phase transition analyses, based on the \textit{one-loop} 
daisy resummed thermal effective potential (c.f. Appendix~\ref{sec:4d-veff}), produce a pressure that is correct to NNLO at $\mathcal{O}(g^3)$
for the broken phase contributions, but are lacking two-loop $\mathcal{O}(g^2)$ terms in the symmetric phase.
Also, such analyses include some $\mathcal{O}(g^4)$ corrections, but are not complete at  $\text{N}^3$LO. 
In our computation in this article at hand, we provide complete accuracy at  $\text{N}^3$LO: this requires a three-loop computation for the hard contributions, and a two-loop computation of the soft contributions and the broken phase effective potential; technical details can be found in Appendix \ref{sec:details}. 
At $\text{N}^4$LO  the $\mathcal{O}(g^5)$ correction would require a three-loop computation in the 3d EFT, which we do not include here, but leave  to future work. It is tempting to pursue this correction, as it is the last term that can still be obtained in perturbation theory, c.f. \cite{Kajantie:2002wa}.

\begin{figure}[t]
\centering
\includegraphics[width=0.49\textwidth]{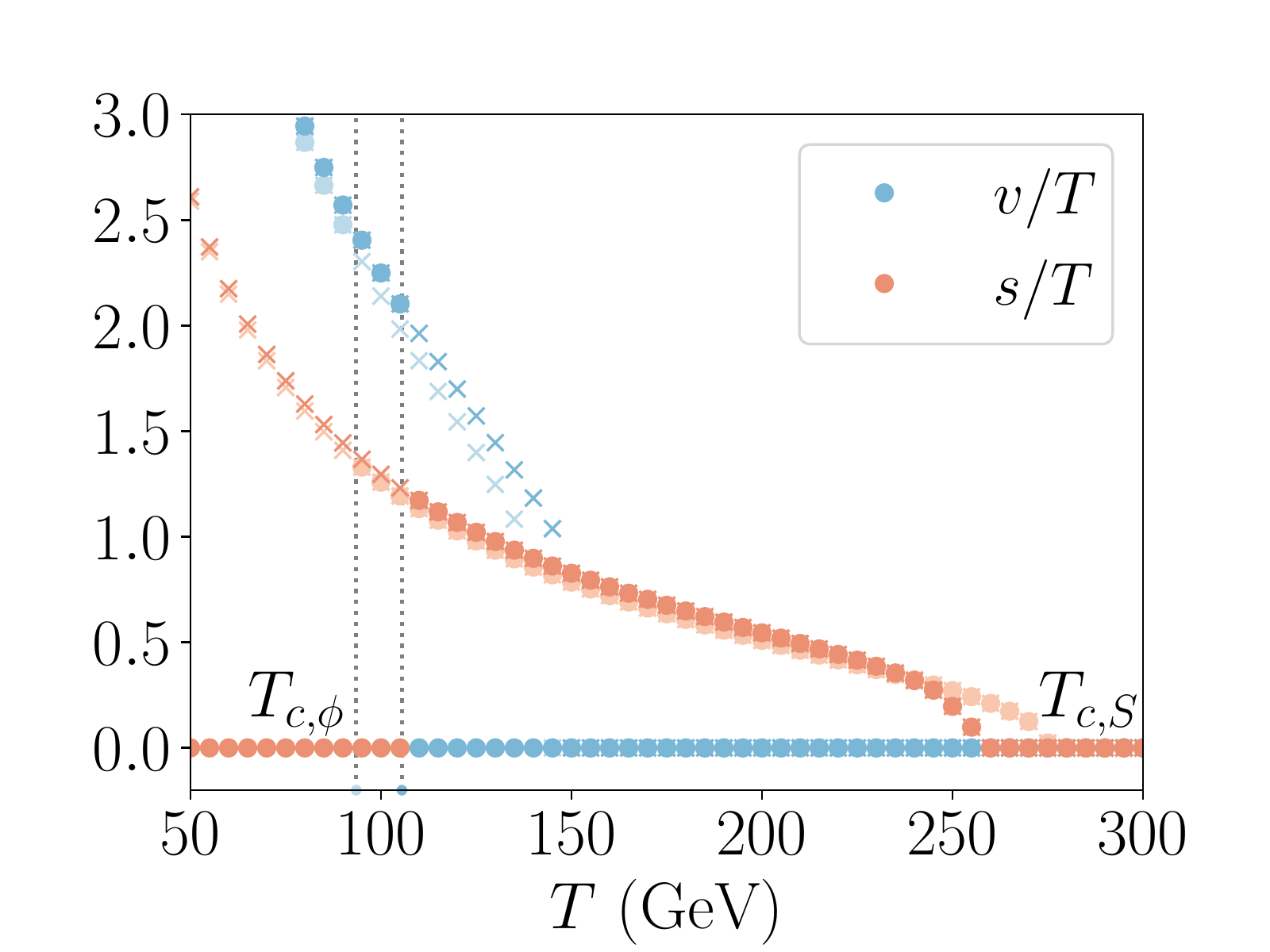} 
\includegraphics[width=0.49\textwidth]{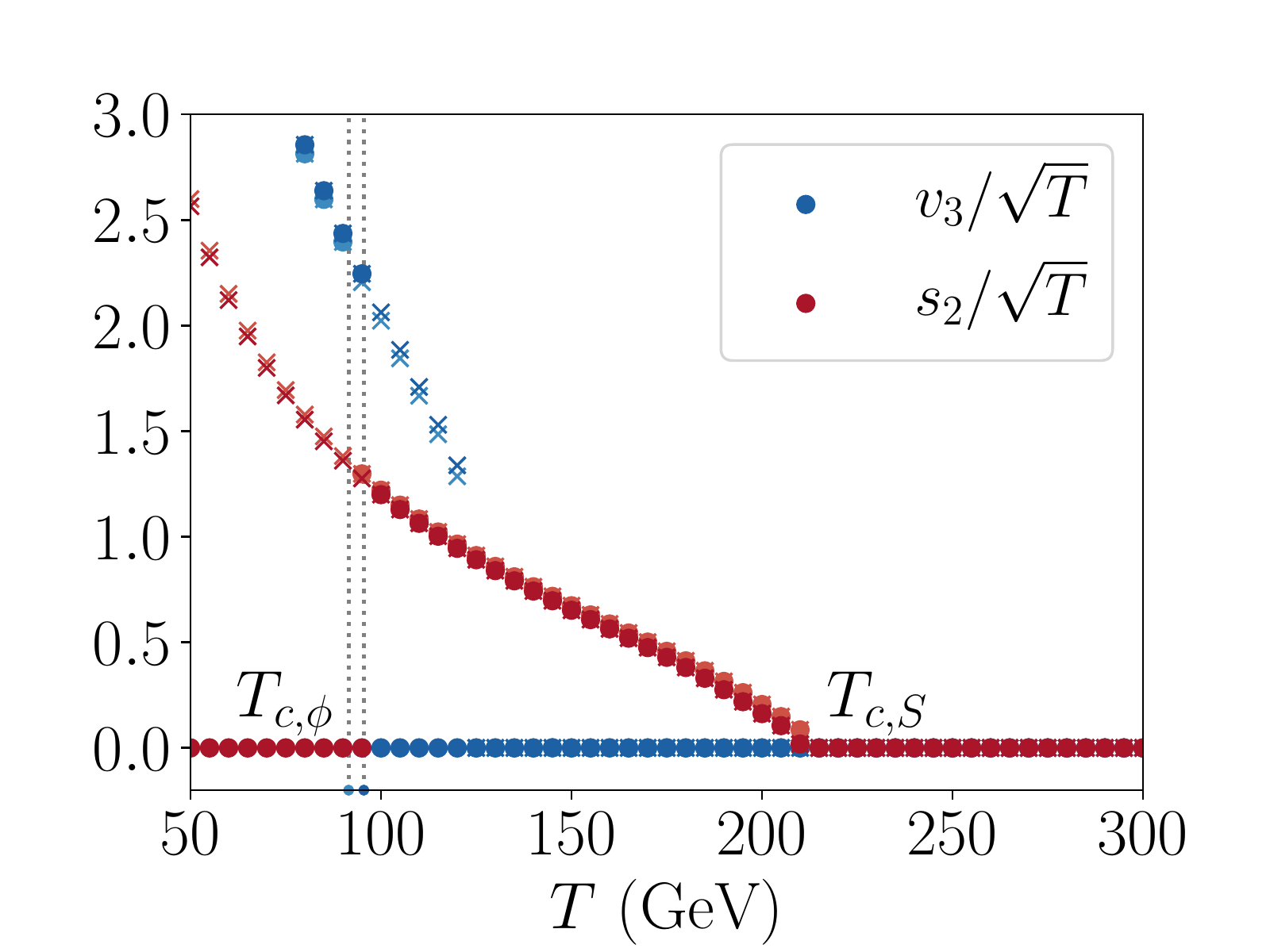} 
\caption{
Minima of the real part of the effective potential as a function of the temperature, determined in a one-loop approximation (left) and at two-loop within the 3d EFT with NLO dimensional reduction (right).
Blue (red) denotes the Higgs (singlet) direction, and dots (crosses) present global (local) minima.
Local minima describe the metastable phases.
The different dark and light colours 
present
two choices for the 4d RG scale $\mu$: $0.5 \pi T$ and  $2 \pi T$. 
The vertical lines depict the critical temperatures for the
second transition ($T_{c,\phi}$), 
and the band between them the theoretical uncertainty in their determination.
We note that the critical temperatures for the first transition ($T_{c,S}$) is very sensitive to two-loop corrections, while $T_{c,\phi}$ is not.
} 
\label{fig:minima}
\end{figure}

\begin{figure}[t]
\centering
\includegraphics[width=0.7\textwidth]{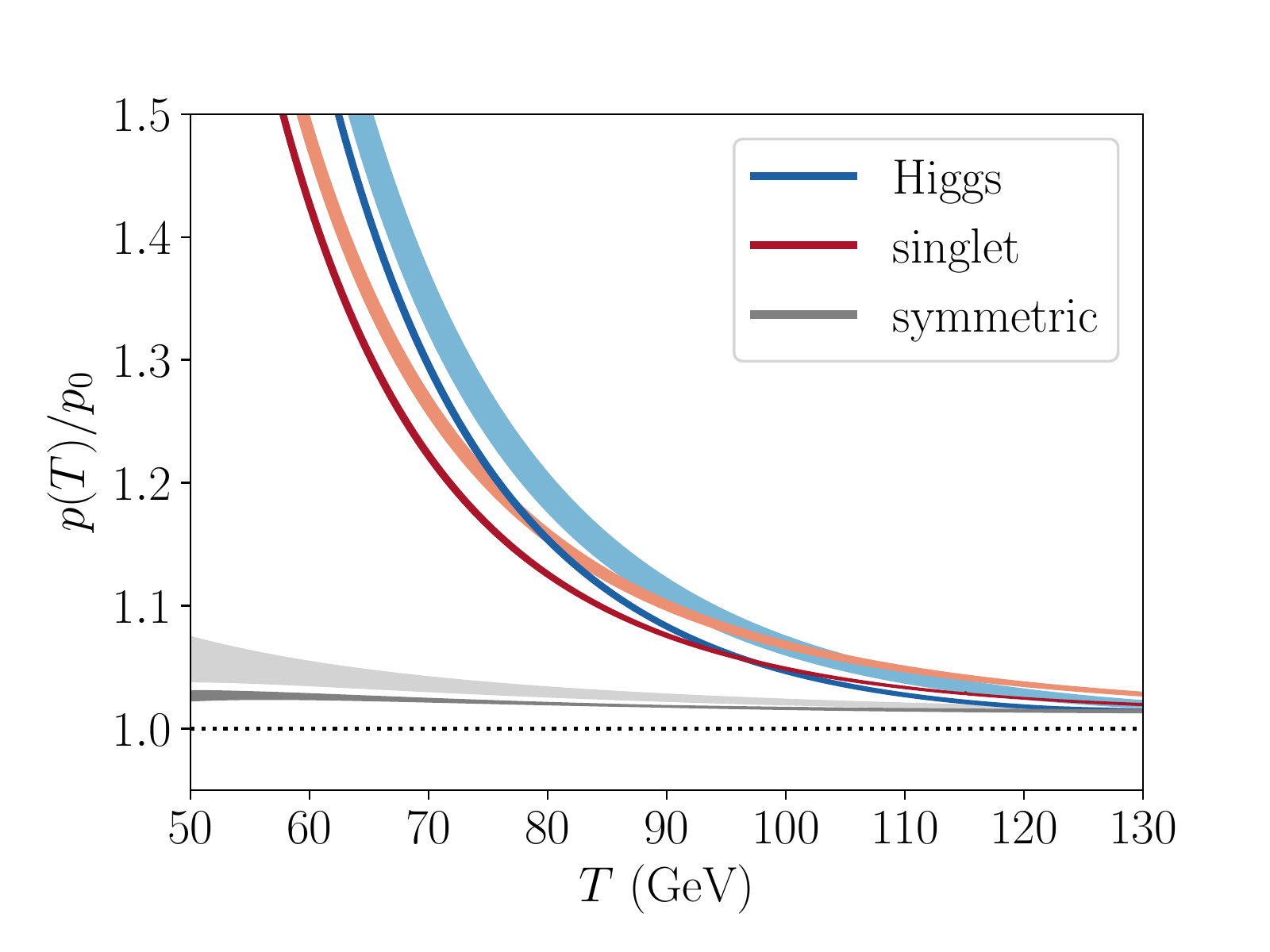} 
\caption{
Pressure of each phase as function of temperature, normalised by the leading order pressure $p_0$. 
Bands correspond to varying the 4d RG scale $\mu$ between $0.5 \pi T$ and  $2 \pi T$.  Darker colours correspond to $\text{N}^3$LO computation, and lighter colours to a mere one-loop approximation.
The result of the $\text{N}^3$LO computation is less sensitive to the RG scale, and corrections to $p_0$ in all phases are mildly smaller than in the mere one-loop approximation. 
} 
\label{fig:pressure}
\end{figure}

\begin{figure}[t]
\centering
\includegraphics[width=0.7\textwidth]{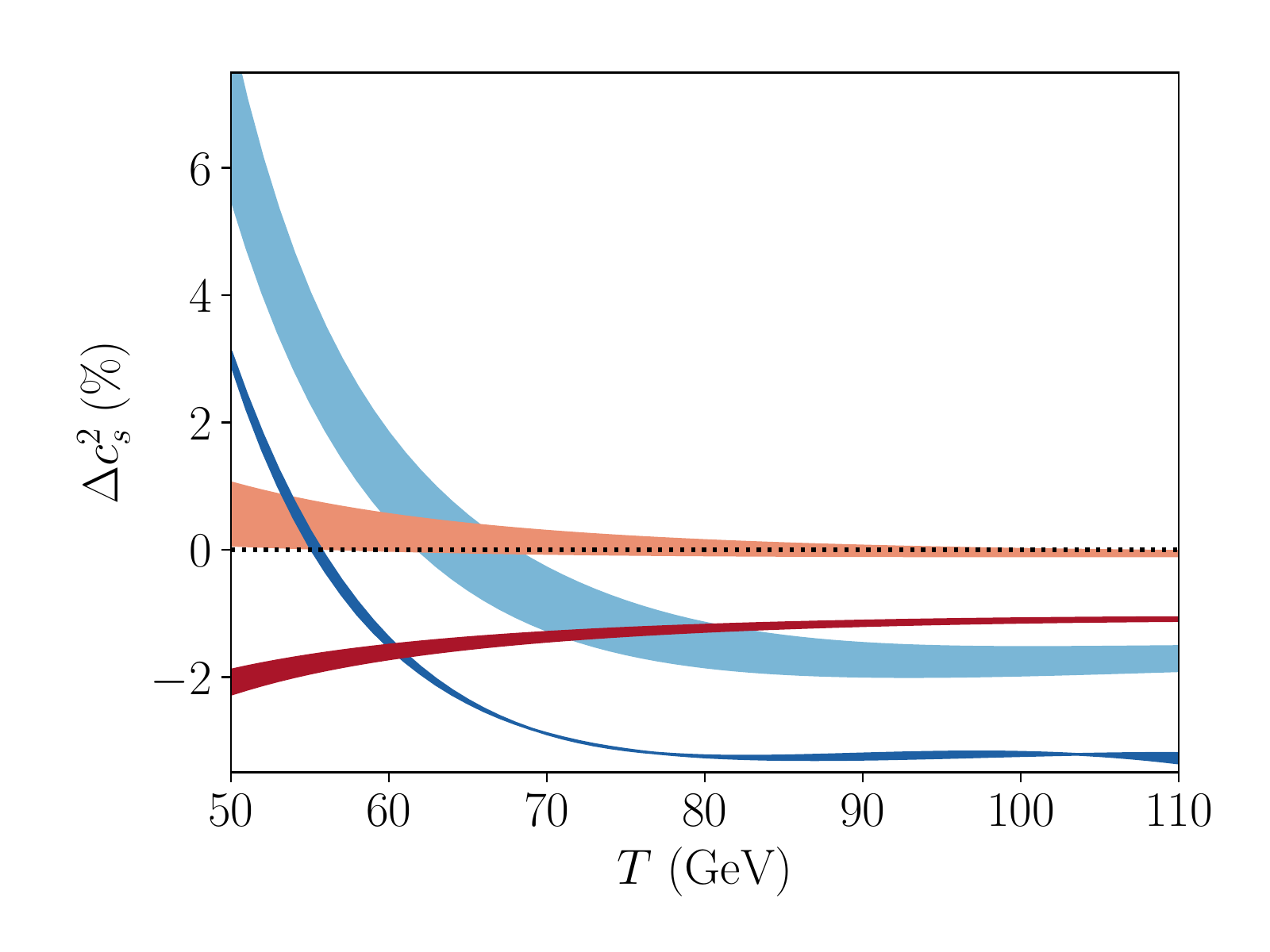} 
\caption{
Percentual deviation in the speed of sound from $c_s^2 = 1/3$ in the singlet phase (red) and Higgs phase (blue) as a function of the temperature. The horizontal line indicates the value of a relativistic gas of particles. Colour codes are the same as in Fig.~\ref{fig:pressure}
} 
\label{fig:sos}
\end{figure}

\paragraph*{Numerical study}
For our numerical analysis, we select a single, representative benchmark point
with a two-step phase transition
\cite{Patel:2012pi,Inoue:2015pza,Blinov:2015sna,Niemi:2020hto,Bell:2020gug}
with a strong second transition to the Higgs phase.
\begin{align}
\text{BM:} \quad (m_S, \lambda_S, \lambda_m) &= (160\, \text{GeV}, 1.0, 1.6),\label{eq:BM}
\end{align}
where $m_S$ is the singlet mass.
   
The order parameters of the
different phases are depicted in Fig.~\ref{fig:minima}.
The Higgs (singlet) background field is denoted by $v$ ($s$).
The pressure and sound speed -- computed from Eq.~\eqref{eq:speedofsound} -- are shown in 
Figs.~\ref{fig:pressure} and \ref{fig:sos}, respectively.
In all these plots, we present a comparison of our full $\text{N}^3$LO computation with the frequently used, sole one-loop approximation \cite{Quiros:1999jp} 
(cf. Appendix~\ref{sec:4d-veff}).  
In addition, we vary the RG scale $\mu$ from $0.5\, \pi T$ to $2\, \pi T$, to monitor convergence, and we find in all the aforementioned figures that in the full $\text{N}^3$LO computation the sensitivity to the 4d theory RG scale is milder, as expected.  
For the 3d RG scales we have used $\bar{\mu}_3=\mu_3 = T$ in Fig.~\ref{fig:minima} and $\bar{\mu}_3=\mu_3 = 1.25 \bar{g}^2_3$ in Figs.~\ref{fig:pressure} and \ref{fig:sos}. 
We note that our use of minima of the effective potential as order parameters leads to an undesired gauge-dependence that signals an
inconsistent perturbative expansion. We carefully comment on this issue in Appendix~\ref{sec:order-parameter}
and argue why this does not compromise our discussion regarding the sound speed.   

We find that the deviations in the speed of sound squared from $c_s^2 = 1/3$ are small in the singlet phase, where they remain subpercent over the entire temperature range for the one-loop result. Deviations are somewhat more apparent in the $\text{N}^3$LO analysis, of the order of $\gtrsim 1\%$. Note that the regime where the deviations are largest is the regime with most supercooling, where the high-temperature 
expansion might not be reliable.
In the Higgs phase, the one-loop result deviates from the relativistic value by $\sim 2\%$ over a large range of temperatures. Again, the deviations in the $\text{N}^3$LO analysis are larger, $\sim 3 - 4 \%$. In both cases, the speed of sound exceeds the relativistic value in the less trustworthy regime of strong supercooling.

%
\section{Impact on gravitational wave predictions}
\label{sec:GWs}

\begin{figure}[t]
\centering  
\includegraphics[width=0.49\textwidth]{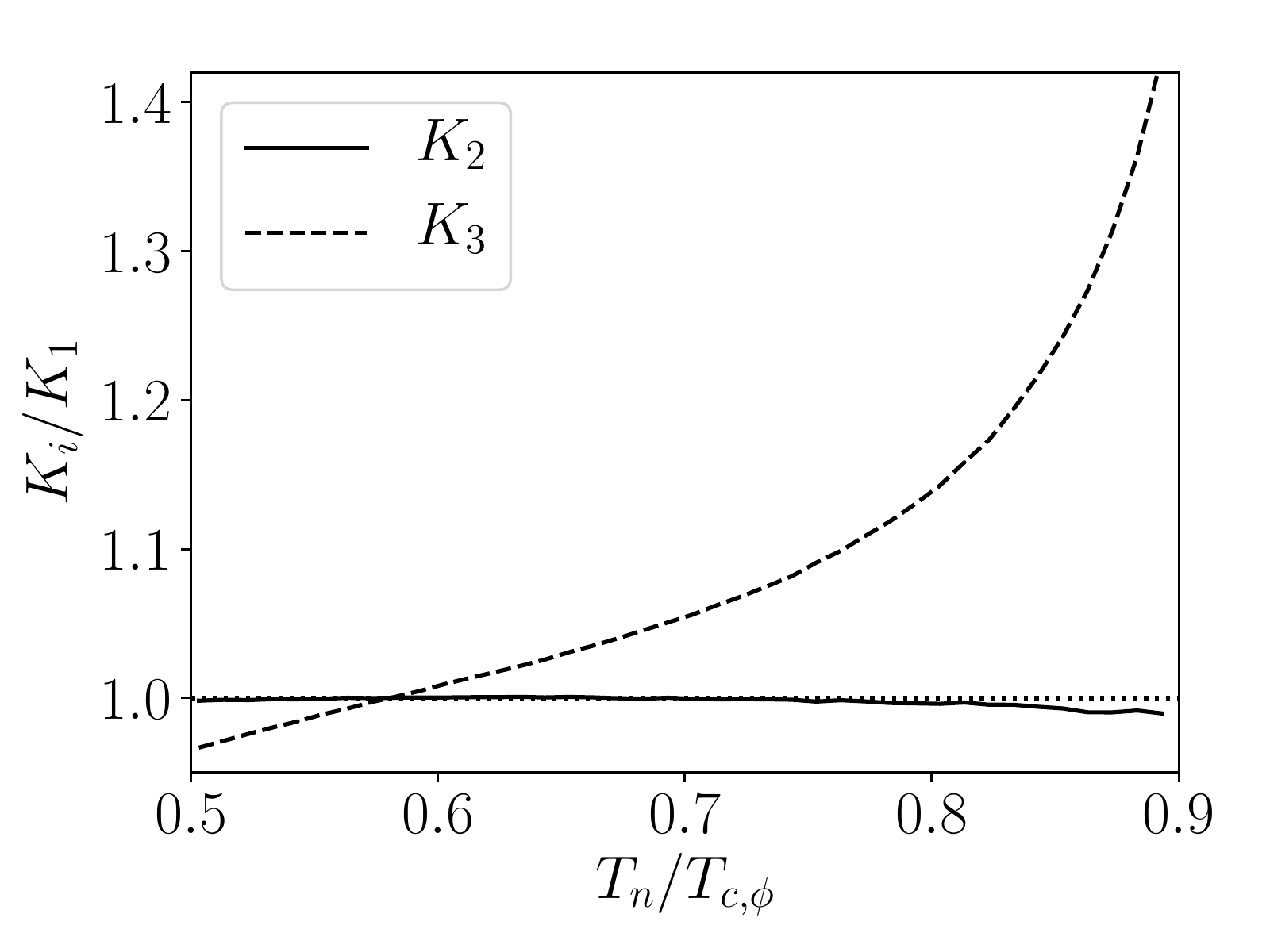} 
\includegraphics[width=0.49\textwidth]{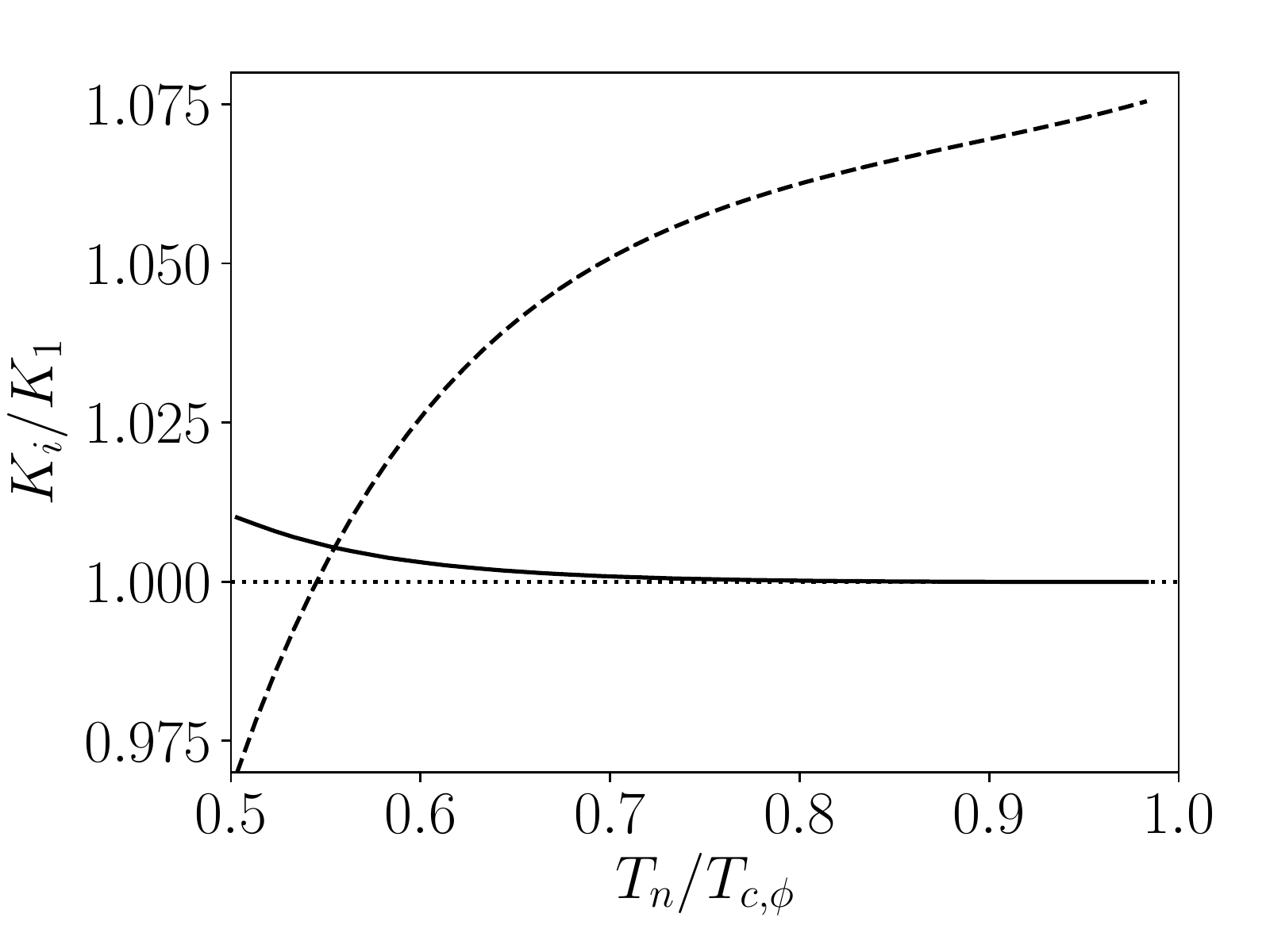}
\caption{
Ratio of the kinetic energy fraction computed in Method 2 and Method 3 to the result of Method 1. 
The left plot shows a hybrid with $v_w = 0.65$ and the right plot a detonation with wall velocity $v_w = 0.99$.}
\label{fig:kratio}
\end{figure}

We have seen in Sec. \ref{sec:gwtheory} that the gravitational wave spectrum depends on the sound speed through the kinetic energy fraction $K$ and through an explicit appearance in Eq.~\eqref{eq:GWfit}. Fig.~\ref{fig:sos} demonstrates that, for $N=1$, the deviations from $c_s = 1/\sqrt 3$ are moderate, so we do not expect major deviations in the gravitational wave spectrum from the explicit dependence on $c_s$. As demonstrated in \cite{Giese:2020znk} however, a small deviation in the sound speed can already lead to a significant effect on the kinetic energy budget. Bear in mind that the gravitational wave amplitude scales with $K^2$ or $K^{3/2}$, which amplifies the dependence. We will quantify the effect of the accurate computation of the sound speed on the gravitational wave spectrum by comparing the kinetic energy budget computed in the bag model (Method 3) with the computations accounting for the deviations in the sound speed (Method 1 and 2). 
In all following plots, we demonstrate the results obtained in the 3d EFT with fixed RG scales $\mu = 1.25 \; \pi T$ and $\mu_3 = \bar{\mu}_3 = 1.25 \; \bar{g}^2_3$. 

It should be noted that an accurate prediction of the gravitational wave spectrum depends on the nucleation temperature, which enters in $H_*,R_*, f_{\rm p,0}$ and $K$. A self-consistent determination of the nucleation temperature requires a computation of the bubble nucleation rate within the 3d nucleation EFT \cite{Gould:2021ccf} (for applications, see \cite{Ekstedt:2021kyx,Hirvonen:2021zej}). 
This goes beyond the scope of this work.
Instead we will present our results for $K$ as a function of an \textit{undetermined} nucleation temperature.

One distinguishes 
three types of hydrodynamic solutions, 
characterised
by the wall velocity: (subsonic) deflagrations, in which the fluid is at rest inside the bubble, and forms a shock wave in front of the bubble wall; (supersonic) detonations, for which the fluid in front of the bubble is at rest and a rarefaction wave forms inside; hybrids, or supersonic deflagrations, which consist of a rarefaction wave and a shock. We treat the wall velocity as an external parameter, but in reality, the computation of its value is challenging. Recently, it has been argued that solutions likely fall into one of two categories: deflagrations and ultrarelativistic detonations, but that no solutions exist between the Jouguet velocity and wall velocities with $\gamma_w \lesssim 10$ \cite{Dorsch:2021nje,Laurent:2022jrs}.

\begin{figure}[t]
\centering
\includegraphics[width=0.7\textwidth]{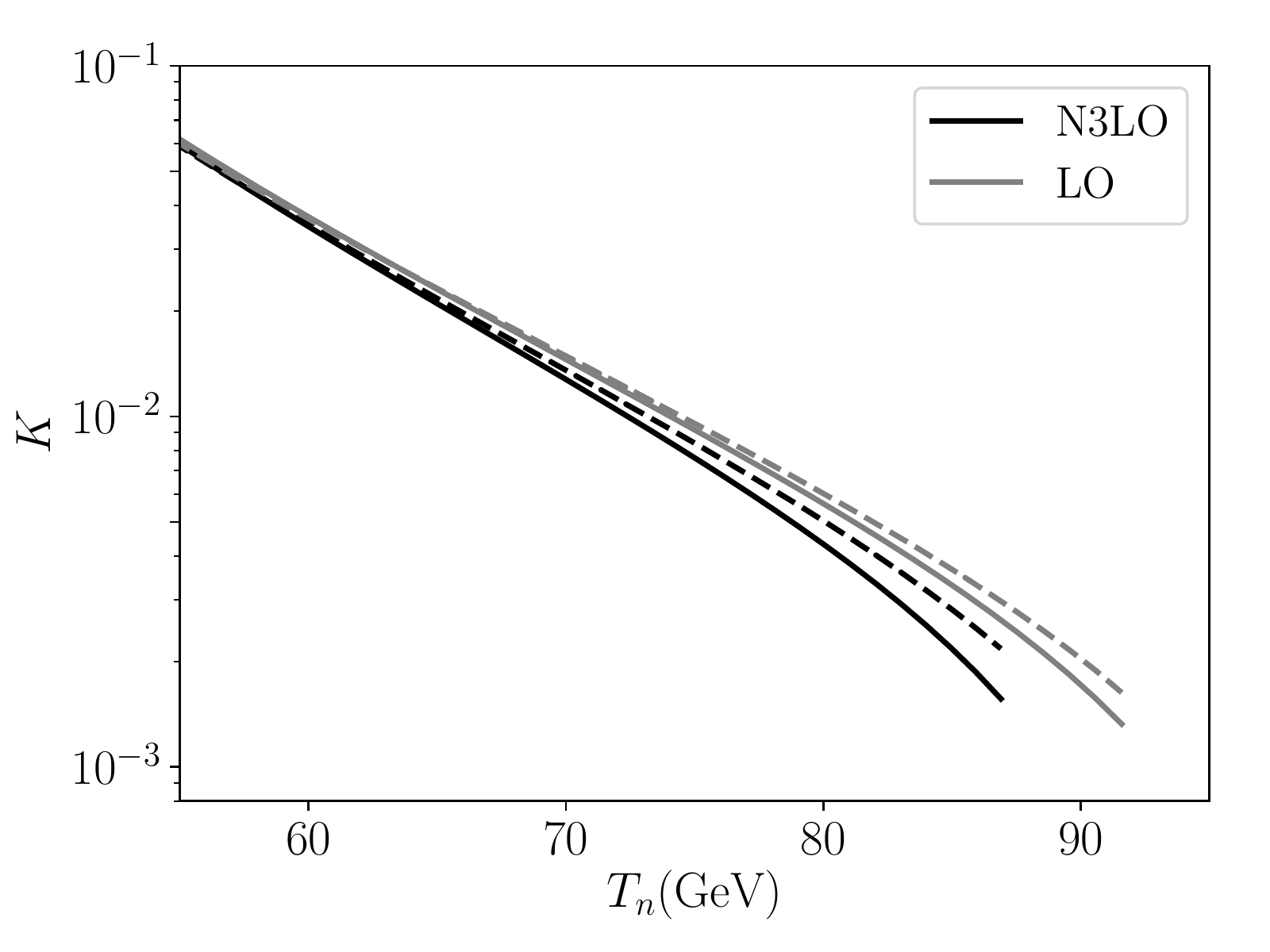} 
\caption{
Absolute value of $K$ as a function of the nucleation temperature for a hybrid ($v_w = 0.65$). The black lines indicate the results of the $\text{N}^3$LO analysis and the gray lines show the result of the one-loop analysis. The dashed lines show the results based on a mapping onto the bag equation of state. 
} 
\label{fig:KofT}
\end{figure}

We have determined the ratio of the kinetic energy fraction $K$ computed in Method 2, $K_2$ and Method 3, $K_3$, to the result in Method 1, $K_1$ for our benchmark point. The results are demonstrated in Fig.~\ref{fig:kratio} for a hybrid and detonation solution. The temperature has been rescaled by the critical temperature of the transition to the singlet to the Higgs phase.
We found that for the deflagration solution, the differences between the three methods are very small ($<\mathcal O(1\%)$) for the entire temperature range.
 This implies that the gravitational wave spectrum can be estimated accurately with $c_s = 1/\sqrt 3$ and a mapping onto the bag equation of state. For a hybrid, the story is different. Especially when the nucleation temperature is relatively close to the critical temperature,\footnote{Note that the upper limit of $T/T_{c,\phi}$ for the hybrid is smaller than unity, since the solutions with $T/T_{c,\phi} \gtrsim 0.9$ and $v_w=0.65$ correspond to detonations.} the difference between $K_3$ and the other methods is more pronounced, approaching $\mathcal O(40\%)$. The estimate based on the bag equation of state typically overestimates the kinetic energy fraction. Over the entire temperature range, $K_2$ reproduces Method 1 very well, meaning that the temperature dependence of the sound speed plays an insignificant role, and a full numerical solution
of the hydrodynamics
is not required. For detonations, the effect of the sound speed is again smaller, with a maximum of $\mathcal O (8\%)$ for $T_n \sim T_{c,\phi}$. For significant supercooling, we see that $K_2$ starts to deviate slightly from $K_1$, implying that the temperature-dependence of the sound speed mildly affects the result.

Fig.~\ref{fig:KofT} showcases the absolute value of $K$ for a hybrid solution as a function of the nucleation temperature,  
for the $\text{N}^3$LO (black) and sole one-loop result (grey).
The solid line displays $K_1$, and the dashed line $K_3$. We see that the difference between $K_1$ and $K_3$ is larger in the $\text{N}^3$LO analysis, which could be expected from 
Fig.~\ref{fig:sos}.
We see that, for a given nucleation temperature, 
the $\text{N}^3$LO computation
yields a smaller value of $K$ than 
the one-loop analysis
(but of course the two approaches do not yield the same nucleation temperature). In addition, our accurate treatment of the speed of sound suppresses the kinetic energy fraction compared to the treatment with $c_s^2=1/3$. For a given temperature, the determination of the pressure 
at one loop
and a determination of $K$ via the bag model together overestimate $K$ by $3\%$ in the strong supercooled regime up to $50\%$ in the regime around $0.9 T_{c,\phi}$.


\section{Speed of sound at a dark sector}
\label{sec:dark-sector}

\begin{figure}[t]
\centering
\includegraphics[width=0.45\textwidth]{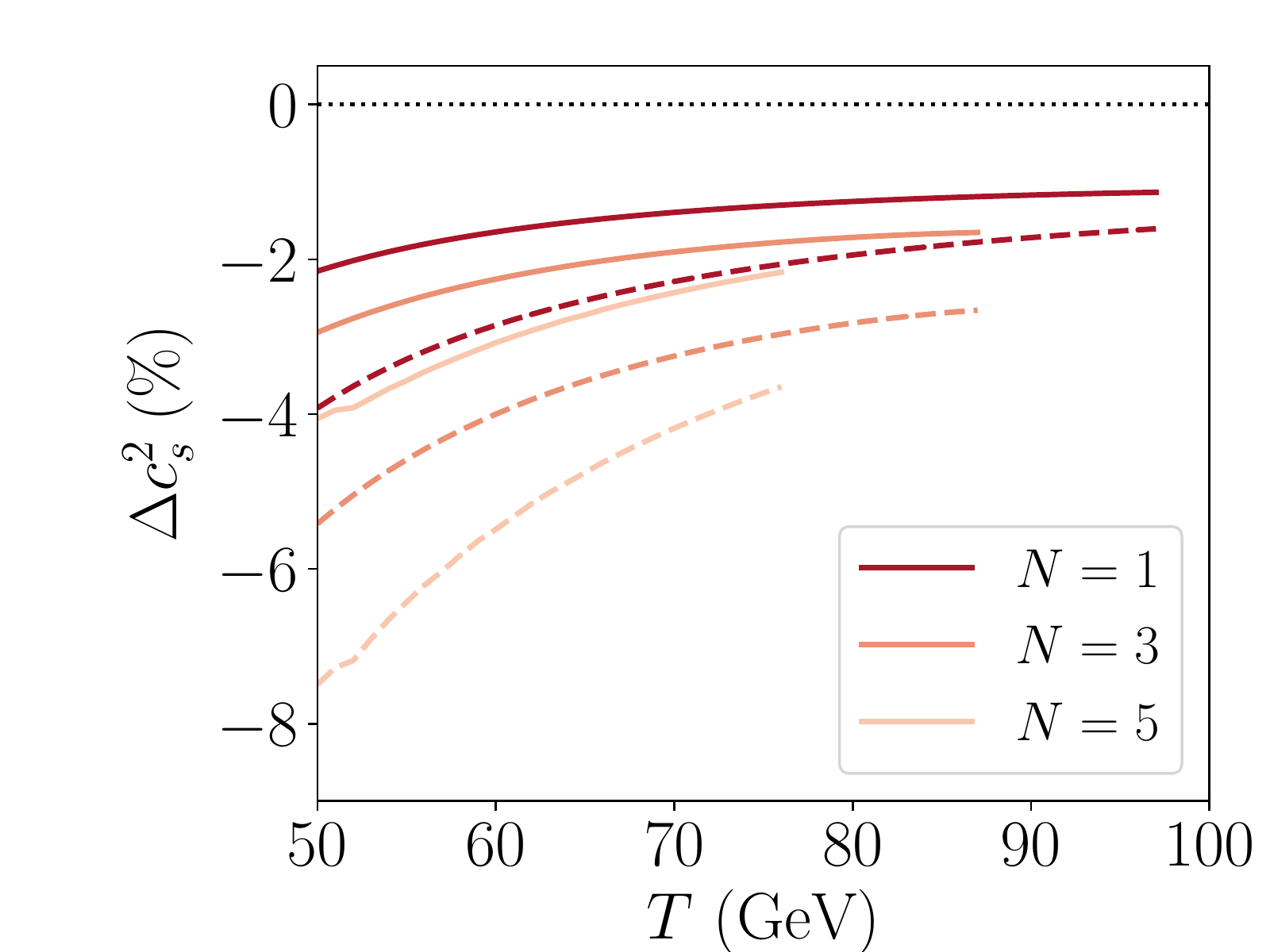} 
\includegraphics[width=0.45\textwidth]{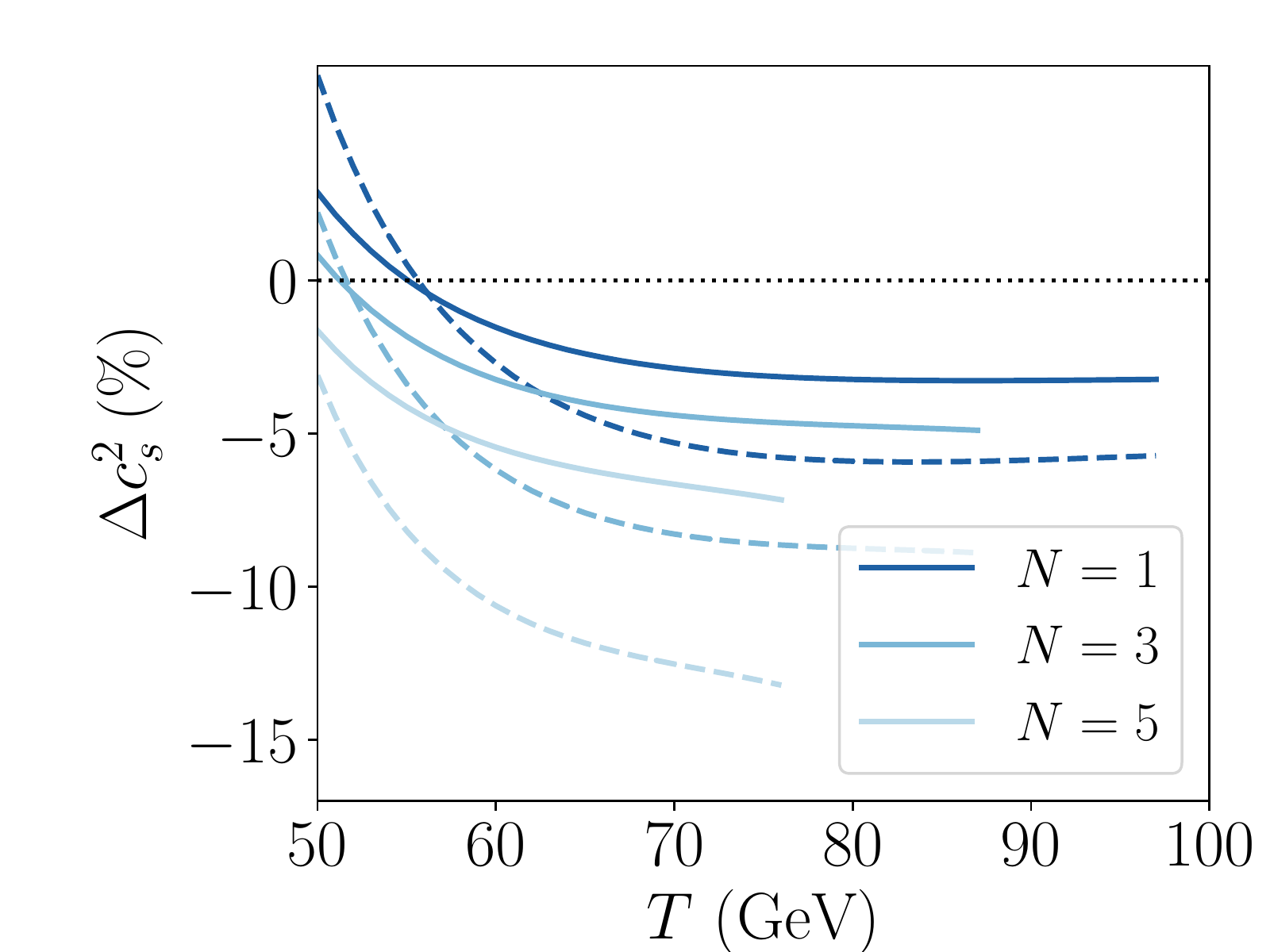}  
\caption{
Sound speed as a function of temperature in the singlet (left) and the Higgs phases (right). The solid lines correspond to the case with 
all fermionic
families
($N_f=3$)
for $N = 1,3,5$,  
and 
dashed lines show 
result with only one fermionic family ($N_f=1$). 
The endpoints of the graphs are set by the critical temperature $T_{c,\phi}$.
} 
\label{fig:soundspeeddark}
\end{figure}

We have seen in the previous section that the sound speed only deviates from $c_s^2 = 1/3$ by a few percent. 
The main reason for this is that the light SM fermions give the dominant contribution to the pressure. 
In this section, we study how the the sound speed behaves when the particle content deviates further from the SM. We investigate two different effects:
\begin{itemize}
	\item{Increasing the number of singlets $N$.}
	\item{Reducing the number of fermions by setting $N_f = 1$. This modification mimics a dark sector with a smaller amount of fermions than the SM.}
\end{itemize}
Fig.~\ref{fig:soundspeeddark} demonstrates the result for the sound speed.
The solid lines show the result for the SM + $N$ singlets, with the usual SM fermionic content. 
As expected, increasing the number of singlets suppresses the sound speed. Again, the deviations are largest in the broken phase, and approach $5\%$. The dotted lines demonstrate the sound speed in the dark sector with only 1/3
of the fermionic content of the SM. We see that this leads to an even stronger suppression in the sound speed, approaching $8\%$ in the singlet phase and $15\%$ in the Higgs phase. This is a significant result, as deviations of that size can suppress the gravitational wave signal by an order of magnitude \cite{Giese:2020znk,Wang:2021dwl} and also affect its shape. 
An even stronger suppression is expected if the particle content gets reduced further.
From this simple exercise
we thus conclude that especially in dark sectors with a particle content that strongly deviates from the SM, a computation of the gravitational wave spectrum requires a careful computation of the sound speed. 
 
%
\section{Discussion}
\label{sec:discussion}

In this work, we have computed the pressure and the corresponding sound speed at $\text{N}^3$LO in formal expansion in $g$ 
(c.f. Eq.~\eqref{eq:pressure-formal} and discussion below it) 
in the Standard Model augmented by $N$ singlets with $\mathcal O(N)$ symmetry. The speed of sound enters in the computation of the energy budget of gravitational waves. Whenever the sound speed deviates from the often-assumed value $c_s^2 = 1/3$, the amplitude and shape of the gravitational wave spectrum are modified. In order to obtain $\text{N}^3$LO accuracy, we used a dimensionally reduced EFT. We worked with a benchmark point defined in Eq.~\eqref{eq:BM} giving rise to a two-step phase transition. We expect our qualitative results to hold for other parameter choices, and other BSM models that accommodate similar strong two-step phase transitions, relevant for observable GW wave signatures.

As known from earlier comparisons between the mere one-loop analysis and higher order analysis in the 3d EFT, thermodynamic quantities such as the critical temperature and latent heat differ significantly between the two computations, potentially leading to multiple order of magnitude differences in the gravitational wave signal. The one-loop result suffers from a disturbing leftover RG-scale dependence, which gets significantly reduced in the result obtained from the 3d EFT. We observe the same reduction in the RG-scale dependence in the pressure and the sound speed. In the $\text{N}^3$LO analysis, suppressions of the sound speed compared to the LO value $c_s^2 = 1/3$ are more significant than in the mere one-loop computation. This suppression affects the energy budget, which sets the amplitude of the gravitational wave signal: in a phase transition with $c_s^2 \neq 1/3$ the energy budget should be computed with the methods of \cite{Giese:2020rtr,Giese:2020znk} instead of the commonly-used bag model. The error associated with using the bag model or an inaccurately computed $c_s^2$ is not as large as the error associated to some of the other parameters of the phase transition\cite{Croon:2020cgk,Gould:2021oba}. 
Nevertheless, especially in the case of hybrid and detonation solutions, the error is non-negligible, and can be as high as $\mathcal O(50\%)$ for the energy budget.

In Sec.~\ref{sec:dark-sector} we further modified the particle content, by increasing the number of singlets, and 
by removing 2/3 of the fermionic content.
These modifications allowed us to envision the possible behaviour of the sound speed in phase transition in dark sectors. We found that both an increase in $N$, as well as the removal of the 
2/3 of the fermions
cause larger deviations in the pressure compared to its LO value. As a result, the sound speed decreases with $N$, and especially small values are obtained in the 
fewer fermion
case. The effect was strongest in the Higgs phase, where the sound speed could decrease by almost $15\%$. Such strong deviations from $c_s^2 = 1/3$ lead to a significant suppression in the gravitational wave signal, by possibly an order of magnitude, and also significantly modify its shape. We thus conclude that computations of the gravitational wave signal from phase transitions in a dark sector require an accurate computation of the sound speed in order to even obtain the right order of magnitude of the signal.

%

\section*{Acknowledgements}

We thank 
Sebastian Bruggisser,
Andreas Ekstedt,
Oliver Gould,
Lauri Niemi,
Philipp Schicho,
Aleksi Vuorinen 
and
Juuso {\"O}sterman
for illuminating discussions
and Mikko Laine for reading and commenting on the manuscript.
TT is supported 
by
National Science Foundation of China grant
no.~19Z103010239. JvdV is supported by the Deutsche Forschungsgemeinschaft under Germany's Excellence Strategy -- EXC 2121 ``Quantum Universe'' -- 390833306, and thanks Nordita for the hospitality during the completion of this work.

%
\appendix
\renewcommand{\thesection}{\Alph{section}}
\renewcommand{\thesubsection}{\Alph{section}.\arabic{subsection}}
\renewcommand{\theequation}{\Alph{section}.\arabic{equation}}


%
\section{Dimensional reduction and pressure with $O(N)$ singlet}
\label{sec:details}

In this appendix we collect multiple details of our computation, in particular the dimensional reduction matching relations for the SM + $O(N)$ singlet, as well as the computation of the pressure. 
For the SM parameters, we use the same power counting in terms of $g$ as in \cite{Brauner:2016fla,Schicho:2021gca}, and in analogy to these references we use
\begin{align}
\mu^2_S \sim (g T)^2, \quad \lambda_S, \lambda_m \sim g^2.
\end{align}
The detailed structure and exact definitions of the 3d EFTs at the soft and ultrasoft scales are discussed in \cite{Brauner:2016fla,Schicho:2021gca,Niemi:2021qvp}. We do not repeat the discussion here, as the only new aspect is the singlet having $N$ components instead of one. Parameters of the 3d EFTs will be denoted with an additional subscript 3, and parameters of the ultrasoft theory with an additional bar.  
We emphasise, that compared to \cite{Gynther:2005av} in which the SM pressure is computed, we use a different power counting for the 3d Higgs mass parameter. In this reference, in the context of pure SM, the proper counting is $\mu^2_{h,3} \sim g^3_3$, which corresponds to a radiatively generated barrier (see for a similar discussion the recent work \cite{Hirvonen:2021zej, Ekstedt:2022zro}), whereas here we use $\mu^2_{h,3} \sim g^2_3$ which corresponds to the case with a tree-level barrier in a two-step phase transition of the Higgs and singlet.  

\subsection{Renormalisation}

We start with 
the zero $T$
renormalisation. 
All the relevant counterterms are listed in \cite{Brauner:2016fla} 
and \cite{Schicho:2021gca}, 
where the case of SM with a single additional singlet was considered. Here, we consider the case of $N$ singlets, and list the counterterms that get modified by the presence of these additional singlets. We use the modified minimal subtraction scheme ($\overline{\rm MS}$).
\begin{align}
	\delta \lambda_h & = \frac{1}{16 \pi^2 \epsilon} \left(\frac{3}{16}(3g^4 + 2 g^2 {g'}^2 + {g'}^4) - 3 g_Y^4 + 12 \lambda_h^2 + \frac N 4 \lambda_m^2 - \frac 1 2 \lambda_h(3 g^2\xi_2 + {g'}^{2}\xi_1) \right)\, ,\\
	\delta \lambda_S &= \frac{1}{16 \pi^2 \epsilon} \left(\lambda_m^2 + \lambda_S^2 (N+8) \right)\, , \\
	\delta \lambda_m &= \frac{1}{16\pi^2 \epsilon} \lambda_m \left(6 \lambda_h + 2 \lambda_m + (N+2)\lambda_S \right)\, , \\
	\delta \mu^2_h &= \frac{1}{16 \pi^2 \epsilon} \left(6 \lambda_h \mu_h^2 + \frac {N}{2} \lambda_m\mu_S^2 \right)\, ,\\
	\delta \mu_S^2 &=\frac{1}{16 \pi^2 \epsilon} \left((N+2)\lambda_S \mu_S^2 + 2\lambda_m \mu_h^2 \right)\, ,
\end{align}
with
$\lambda_h$ the Higgs quartic coupling and $\mu_h$ its mass parameter,
 $g$ and $g'$ the SU(2) and U(1) gauge couplings, $g_Y$ the Yukawa coupling and $\xi_2$ and $\xi_1$ the SU(2) and U(1) gauge fixing parameters, in general covariant gauge, respectively.

The one-loop RG-equations, which encode the running of parameters as a function of the RG-scale $\mu$ to order $\mathcal{O}(g^4)$,  read
\begin{align}
\mu \frac{d}{d\mu} \lambda_{h} &= \Big( \mu \frac{d}{d\mu} \lambda_{h} \Big)_{\text{SM}} + \frac{1}{(4\pi)^2}\Big( \frac{1}{2}N \lambda^2_m \Big) , \\
\mu \frac{d}{d\mu} \mu^2_{h} &= \Big( \mu \frac{d}{d\mu} \mu^2_{h} \Big)_{\text{SM}} +  \frac{1}{(4\pi)^2}\Big(  N \lambda_m \mu^2_S \Big) , \\
\mu \frac{d}{d\mu} \lambda_{S} &=  \frac{1}{(4\pi)^2}\Big(  2 \lambda^2_m + 2(N+8)\lambda^2_S\Big) , \\
\mu \frac{d}{d\mu} \lambda_{m} &=  \frac{2}{(4\pi)^2}  \lambda_m \Big( -\frac{3}{4}(3g^2 + {g'}^2) + 3 g^2_Y + 2 \lambda_m + 6 \lambda_h + (N+2) \lambda_S  \Big) , \\
\mu \frac{d}{d\mu} \mu^2_{S} &=  \frac{1}{(4\pi)^2}\Big( 2 (N+2) \lambda_S \mu^2_S + 4 \lambda_m \mu^2_h \Big). 
\end{align}
For $N=1$, they agree with the RG-equations of \cite{Brauner:2016fla}, but be aware of the different convention for the sign of $\mu_h^2$ and $\mu_S^2$. 

All \MSbar{} parameters in the Lagrangian can be related to pole masses and other physical parameters as in \cite{Kajantie:1995dw,Gould:2019qek,Niemi:2021qvp}, but in this work at hand, we merely use leading, tree-level relations without one-loop corrections. 

\subsection{Dimensional reduction}

For the dimensional reduction procedure,
the Feynman rules of the symmetric phase are listed in \cite{Brauner:2016fla}. Here we only list the Feynman rules for the singlet. The singlet propagator reads 
\begin{align}
\langle S_a(P) S_b(Q) \rangle = \delta_{ab}  \frac{\deltabar(P+Q)}{P^2 + \mu^2_S},
\end{align}
where $P,Q$ denote Euclidean four-momentum 
and $\,\deltabar(K)\equiv T^{-1}\delta_{K_{0},0}(2\pi)^{d}\delta^{(d)}(\vec{k})$ where $\delta_{K_0,0}$
is the Kronecker delta
for vanishing zero component. 
The vertex Feynman rules read
\begin{align}
V(S_a S_b S_c S_d) &= -2 \lambda_S (\delta_{ab} \delta_{cd} + \delta_{ac} \delta_{bd} + \delta_{ad} \delta_{bc}), \\
V(S_a S_b \phi^\dagger_i \phi_j) &= -\lambda_m \delta_{ij} \delta_{ab},
\end{align}
where the indices $a,b,c,d$ label singlet components, for which
$\delta_{aa} = N$, while $i,j$ are fundamental SU(2) indices for the Higgs. 
The 3d matching relations, or short distance coefficients, can be obtained following \cite{Kajantie:1995dw,Braaten:1995cm,Brauner:2016fla,Gould:2021dzl,Schicho:2021gca,Niemi:2021qvp}.%
\footnote{
At the time of our computation, awakening of {\tt DRalgo} \cite{Ekstedt:2022bff} in its lair was still in the future. 
}
We list the matching relations that involve the singlet fields here at 
NLO, i.e. at $\mathcal{O}(g^4)$: 
\begin{align}
\lambda_{h,3} =& \Big(\lambda_{h,3}\Big)_{\text{SM}} + \frac{1}{(4\pi)^2} L_b(\mu) \Big(
    - \frac{1}{4} N \lambda_{m}^{2}
    \Big) , \\
\mu^2_{h,3}(\mu_3) =& \Big( \mu^2_{h,3}(\mu_3) \Big)_{\text{SM}} + \frac{T^2}{24} N \lambda_{m}^{ }(\mu) 
+ \epsilon \; T^2 \beta_{\lambda_m 2} \lambda_m  - \frac{L_b(\mu)}{(4\pi)^2} \Big(
      \frac{1}{2} N \lambda_{m}^{ } \mu_{S}^{2}
  \Big) \nonumber \\
&  
  + \frac{1}{(4\pi)^2}
    \bigg(\frac{3}{4}(3g^2 + \gp^2) L_b(\mu) - 3 \gY^{2} L_f \bigg)
    \bigg(\frac{T^2}{24} N \lambda_m^{ } \bigg)   - \frac{1}{(4\pi)^2} \frac{1}{2} N \lambda^2_{m,3} \Big(c + \ln\Big(\frac{3T}{\mu_3}\Big)\Big)
  \nn &
  - \frac{T^2}{(4\pi)^2} L_b(\mu) N \lambda_{m}^{ } \Big(
      \frac{1}{4}\lambda_h^{ }
    + \frac{5}{24}\lambda_m^{ }
    + \frac{(N+2)}{24}\lambda_S^{ }
  \Big) 
, \\
\lambda_{S,3} =& T \bigg[
    \lambda_S^{ }(\mu)
  - \frac{1}{(4\pi)^2} L_b(\mu) \Big(\lambda_{m}^{2} + (N+8) \lambda_{\sigma}^{2} \Big)
  \bigg] , \\
\lambda_{m,3} =& T \bigg[
  \lambda_{m}^{ }(\mu)
  + \frac{\lambda_m^{ }}{(4\pi)^2} \bigg(
      L_b(\mu)  \Big(
        \frac{3}{4}(3g^2 + \gp^2)
      - 6\lambda_h^{ }
      - 2\lambda_m^{ }
      - (N+2)\lambda_S^{ }
    \Big)
    - 3 L_f(\mu) \gY^{2}
  \bigg) \bigg], \\
\mu^2_{S,3}(\mu_3) =& \mu_{S}^{2}(\mu)
  + T^2 \Big(
      \frac{1}{6}\lambda_m^{ }(\mu)
    + \frac{1}{12}(N+2)\lambda_S^{ }(\mu)
  \Big) + \epsilon \; T^2 \Big( \gamma_{\lambda_m 2} \lambda_m + \gamma_{\lambda_s 2} \lambda_s  \Big) \nonumber \\
& 
- \frac{L_b(\mu)}{(4\pi)^2}\Big(
     2 \lambda_m^{ } \mu_{h}^{2}
    + (N+2) \lambda_S^{ } \mu_{S}^{2}
  \Big)
  \nn &
  + \frac{1}{(4\pi)^2} \Big(
    (3g^2_3 + \gp^2_3) \lambda_{m,3}^{ }
    - 2\lambda_{m,3}^{2}
    - 2(N+2)\lambda_{S,3}^{2}
  \Big) \Big( c + \ln\Big(\frac{3T}{\Lamd} \Big) \Big)
  \nn &
  + \frac{T^2}{(4\pi)^2} \bigg[
      \frac{(2+3L_b(\mu))}{24}(3g^2 + \gp^2) \lambda_m^{ }
- \frac{1}{4}(3L_b(\mu) - L_f(\mu)) \gY^{2}\lambda_{m}^{ } 
\nn &
    - L_b(\mu) \bigg(
      \Big(
          \lambda_h^{ }
        + \frac{1}{12}(N+6)\lambda_m^{ }
        + \frac{1}{6}(N+2) \lambda_S^{ }
      \Big) \lambda_m^{ }
    + \frac{1}{12}(N+2)(N+8) \lambda_{S}^{2} \bigg)
  \bigg],
\end{align}
where $\mu_3$ denotes the RG-scale of the 3d EFT.
Here we have used the shorthand notation
\begin{align}
c &\equiv \frac 1 2 \left( \ln{\left(\frac{8\pi}{9}\right)} + \frac{\zeta'(2)}{\zeta(2)} -2\gamma \right), \\
L_b &\equiv 2\left( \ln{\frac{\mu}{4\pi T}} + \gamma \right), \qquad \qquad \qquad L_f \equiv L_b + 4\ln 2, \\
I_b &\equiv  \bigg( 1 + \ln\Big(\frac{\mu}{4\pi T}\Big) + \frac{\zeta'(-1)}{\zeta(-1)} \bigg). \label{eq:lblf}
\end{align}
For the mass parameter we have included the $\mathcal{O}(\epsilon \; T^2 g^2)$ level pieces, as these contribute to the ultrasoft pressure $p_{\text{M2}}$ when multiplied with $1/\epsilon$ poles therein. The coefficients $\beta_{\lambda_m 2}, \gamma_{\lambda_m 2}$ and $\gamma_{\lambda_s 2}$ read
\begin{align}
\beta_{\lambda_m 2} &= \frac{N}{12} I_b , \qquad 
\gamma_{\lambda_m 2} = \frac{1}{3} I_b ,  \qquad
\gamma_{\lambda_s 2} = \frac{N+2}{6} I_b. 
\end{align}
Similar pure SM contributions to the Higgs mass parameter can be found in \cite{Gynther:2005av} in Appendix B, there denoted by $\beta_{A 2},\beta_{B 2},\beta_{\lambda 2},\beta_{Y 2}$.
The 3d gauge parameters do not receive any singlet contributions at NLO 
and can be read off from \cite{Kajantie:1995dw,Brauner:2016fla}.  
The Debye masses for the temporal scalars read 
\begin{align}
m^2_D &= \Big( m^2_D \Big)_{\text{SM}} + \frac{N}{(4\pi)^2} \frac{T^2}{12} g^2 \lambda_m , \label{eq:Debye1} \\
{m}'^2_D &= \Big( {m}'^2_D \Big)_{\text{SM}} + \frac{N}{(4\pi)^2} \frac{T^2}{12} {g'}^2 \lambda_m \label{eq:Debye2}. 
\end{align}
The SM parts at two-loop order can be found in Refs. \cite{Gynther:2005dj,Gynther:2005av}. 
The singlet interacts with the temporal scalar fields through \cite{Schicho:2021gca}
\begin{align}
y_3 &= T \frac{1}{(4\pi)^2} \frac{1}{2} g^2 \lambda_m , \\
y_3' &= T \frac{1}{(4\pi)^2} \frac{1}{2} {g'}^2 \lambda_m  . 
\end{align}
However, since these interactions only start at $\mathcal{O}(g^4)$, they do not contribute to any of our further expressions.

\subsection{Coefficient of the unit operator, or pressure, in the symmetric phase}
\label{sec:p-sym}

In addition to the matching of the couplings and masses, we also need the matching of the unit operator (c.f. \cite{Braaten:1995cm} Sec. 3A), which amounts to computing vacuum contributions to the pressure from the hard and soft modes, as in \cite{Gynther:2005dj}. Note that by convention, we include the similar ultrasoft pieces of the symmetric phase in the effective potential. 

\paragraph{Leading order pressure}

At leading order, the pressure is given in terms of the sum-integrals \cite{Braaten:1995jr}
\begin{align}
\mathcal{I}'_0 &\equiv -\sumint{P} \ln(P^2) = \frac{\pi^2}{45} T^4 + \mathcal{O}(\epsilon) , \\
\widetilde{\mathcal{I}}'_0 &\equiv -\sumint{\{P\} } \ln(P^2) = \frac{7}{8} \frac{\pi^2}{45} T^4 + \mathcal{O}(\epsilon),
\end{align} 
arising from hard mode contributions of one-loop vacuum bubble diagrams. For the definition of the sum-integration measure, see \cite{Braaten:1995jr}. 
Following \cite{Gynther:2005dj}, different fields contribute to the total pressure as 
\begin{align}
\text{(fermions)} &= -\Big(\underbrace{1}_{\text{lepton singlets}} + \underbrace{2 N_c}_{\text{quark singlets}} \nonumber \\
&+ d_F ( \underbrace{1}_{\text{lepton doublets}} + \underbrace{N_c}_{\text{quark doublets}}   ) \Big) N_f
\; \widetilde{\mathcal{I}}'_0 , \\
\text{(gauge fields)} &= \frac{1}{2} D \Big( \underbrace{d_A}_{\text{weak bosons}} + \underbrace{1}_{\text{photon}}  + \underbrace{(N^2_c-1)}_{\text{gluons}}  \Big)  \mathcal{I}'_0, \\
\text{(ghosts)} &= -\Big( \underbrace{d_A}_{\text{weak bosons}} + \underbrace{1}_{\text{photon}}  + \underbrace{(N^2_c-1)}_{\text{gluons}}  \Big) \mathcal{I}'_0, \\
\text{(Higgs)} &= d_F \mathcal{I}'_0, \\
\text{(singlet)} &= \frac{N}{2} \mathcal{I}'_0. 
\end{align} 
To label different contributions, we have used the following notation of \cite{Gynther:2005dj} for the constants: $N_f=3$ for the number of fermion families, $N_c = 3$ for the SU(3) colour group, $d_F = 2$ for the dimensionality of the fundamental representation of SU(2) and $d_A=3$ the dimensionality of the adjoint representation of SU(2). In total, 
\begin{align}
p^{\text{LO}}(T) &= \frac{\pi^2}{90} T^4 \bigg(1 + d_A + (N^2_c-1) + d_F + N + \frac{7}{8}N_f \Big(1 + d_F + N_c (d_F + 2) \Big) \bigg) \nonumber \\ 
&= (28 + N + 26.25 N_f)\frac{\pi^2}{90} T^4.
\end{align}
This is the leading order pressure in the symmetric phase.

\paragraph{Higher order corrections}
The 
full
pressure in the symmetric phase has the form
\begin{align}
p_{\text{sym}}(T) = p_{\text{E}}(T) + p_{\text{M1}}(T) + p_{\text{QCD}}(T) + \mathcal{O}(g^{5} T^4).   \label{eq:psym}
\end{align}
Here, different contributions are organised in a manner similar to \cite{Gynther:2005dj,Gynther:2005av}: $p_{\text{E}}$ collects the hard mode contributions, and $p_{\text{M1}}$ the soft mode contributions from temporal scalars. The subscripts denote electric and magnetic contributions, as in hot EQCD \cite{Braaten:1995jr}. We note that by convention, the leading order QCD contributions are included in $p_{\text{E}}$, while higher order QCD corrections are collected in $p_{\text{QCD}}$ 
\cite{Braaten:1995jr,Arnold:1994ps,Arnold:1994eb,Zhai:1995ac,Kajantie:2002wa}.
We do not include these QCD corrections in our computation,
as consecutive orders are known to fluctuate around the ideal gas pressure
-- due to the large values of the strong coupling $g_s$ and $g_Y$ --  unless the temperature is asymptotically large, 
and here our focus is solely on the EW corrections.

In addition, for $p_{\text{sym}}$ there would also be an ultrasoft, or additional magnetic contribution  
$p_{\text{M2}}(T)$ \cite{Gynther:2005dj,Gynther:2005av} that arises from one- and two-loop diagrams with ultrasoft scalars. However, by convention we include these contributions in the effective potential part of the pressure, i.e. $p_{\text{M2}}(T) = -T V^{\text{3d}}_{\text{eff}}(0,0)$ (c.f. Sec.~\ref{sec:veff}), 
where both Higgs and singlet background fields vanish in the symmetric phase.

To reach the $\text{N}^3$LO, or $\mathcal{O}(g^4)$ precision for the vacuum diagrams requires a computation up to -- and including -- three-loop topologies, for the hard scale contributions. 
These can be split into
\begin{align}
p_{\text{E}}(T) = p^{\text{SM}}_{\text{E}}(T) + p^{\text{singlet}}_{\text{E}}(T),   
\end{align}
where the SM contribution can be read off from Eq.~(20) and Appendix A in \cite{Gynther:2005dj}.%
\footnote{
We note that Ref.~\cite{Laine:2015kra} presents minor corrections to the results of \cite{Gynther:2005dj,Gynther:2005av}, and for the parts relevant for our computation, we have confirmed and used the results of \cite{Laine:2015kra}.
}
In this work, we include the singlet contributions
\begin{align}
p^{\text{singlet}}_{\text{E}}(T) &= T^4 \bigg[ \alpha^{\text{singlet}}_{E1} +  \lambda_m(\mu) \alpha_{E \lambda_m} +  \lambda_S(\mu) \alpha_{E \lambda_S} \nonumber \\
& + \frac{1}{(4\pi)^2} \bigg( g^2 \lambda_m \alpha_{EA \lambda_m} + {g'}^2 \lambda_m \alpha_{EB \lambda_m} + g^2_Y \lambda_m \alpha_{EY \lambda_m} \nonumber \\
& + \lambda^2_S \alpha_{E \lambda_S \lambda_S} + \lambda^2_m \alpha_{E \lambda_m \lambda_m} + \lambda_S \lambda_m \alpha_{E \lambda_S \lambda_m} + \lambda_h \lambda_m \alpha_{E \lambda_h \lambda_m} \bigg)  \bigg] \nonumber \\
& + T^2 \bigg[ \mu^2_S(\mu) \alpha_{E \mu^2_S} + \frac{1}{(4\pi)^2} \bigg( \lambda_m \mu^2_S  \alpha_{E \lambda_m \mu^2_S} + \lambda_m \mu^2_h  \alpha_{E \lambda_m \mu^2_h} + \lambda_S \mu^2_S  \alpha_{E \lambda_S \mu^2_S} \bigg) \bigg] \nonumber \\
&  + \frac{\mu^4_S}{(4\pi)^2} \alpha_{E \mu^4_S} + \mathcal{O}(g^6),    \label{eq:psinglet}
\end{align}
where we used notation similar to \cite{Gynther:2005dj,Gynther:2005av}.
We list the results for all coefficients $\alpha$ in Eqs.~\eqref{eq:alphaE1}-\eqref{eq:alphaEmus4}.

The soft contribution from the temporal scalars reads
\begin{align}
p_{\text{M1}}(T) = p^{\text{SM}}_{\text{M1}}(T),
\end{align}
i.e. it is of exactly the same functional form as in the SM, see Eq.~(12) in \cite{Gynther:2005av}. Note, however, that the singlet contributes to the two-loop Debye masses $m_D$ and $m_D'$ as in Eqs.~\eqref{eq:Debye1} and \eqref{eq:Debye2}, and hence to 
$p_{\text{M1}}$.
Note that three-loop diagrams contribute to $p_{\text{M1}}$ at  $\text{N}^4$LO, or $\mathcal{O}(g^5)$, but we include them nonetheless, since they provide the \textit{complete soft contribution} at $\mathcal{O}(g^5)$.
There are three-loop diagrams involving the singlet and temporal scalars, but since their respective couplings are loop-induced, they are relatively suppressed and do not contribute at $\mathcal{O}(g^5)$.

The ultrasoft contribution to the pressure reads
\begin{align}
p_{\text{M2}}(T) = p^{\text{SM}}_{\text{M2}}(T) + p^{\text{singlet}}_{\text{M2}}(T),
\end{align}
where the SM piece can be read off from Eq.~(15) in \cite{Gynther:2005av} and
\begin{align}
p^{\text{singlet}}_{\text{M2}}(T) &= T \bigg(\frac{N}{12 \pi} \bar{\mu}^3_{s,3} - \frac{N}{(4\pi)^2} \Big( \frac{1}{4} (N+2) \bar{\lambda}_{s,3} \bar{\mu}^2_{s,3} +  \bar{\lambda}_{m,3} \bar{\mu}_{s,3} \bar{\mu}_{h,3}   \Big) \bigg) .
\end{align}
As in \cite{Gynther:2005av}, we compute $p_{\text{M2}}(T)$ at two-loop order, providing $\mathcal{O}(g^4)$ accuracy. 

Finally, we comment that the divergent $1/\epsilon$ poles in $p_{\text{E}}$, $p_{\text{M1}}$ and $p_{\text{M2}}$ cancel
when all these terms are summed together, i.e. the cancellation happens between contributions from different scales.%
\footnote{
Here lies a technical subtlety. All $1/\epsilon$ poles indeed cancel, but this cancellation happens exactly only when the ultrasoft pieces are expanded to $\mathcal{O}(g^4)$. As we have explained in Sec.~\ref{sec:pressure}, we do not expand the ultrasoft pieces, but in practice we remove $1/\epsilon$ terms manually.        
} 
Note that $p^{\text{singlet}}_{\text{M2}}$ is finite 
and $p_{\text{M1}}$ does not have divergences related to the singlet contributions. Therefore all leftover divergent singlet pieces from the hard modes are cancelled by the ultrasoft Higgs-gauge field contribution that is proportional to $\bar{\mu}^2_{h,3}$ and that encodes the contribution from the singlet portal coupling.
    
\paragraph{Vacuum Feynman diagrams at the hard scale with the singlet}

\begin{figure}[t]
\centering
\includegraphics[width=0.85\textwidth]{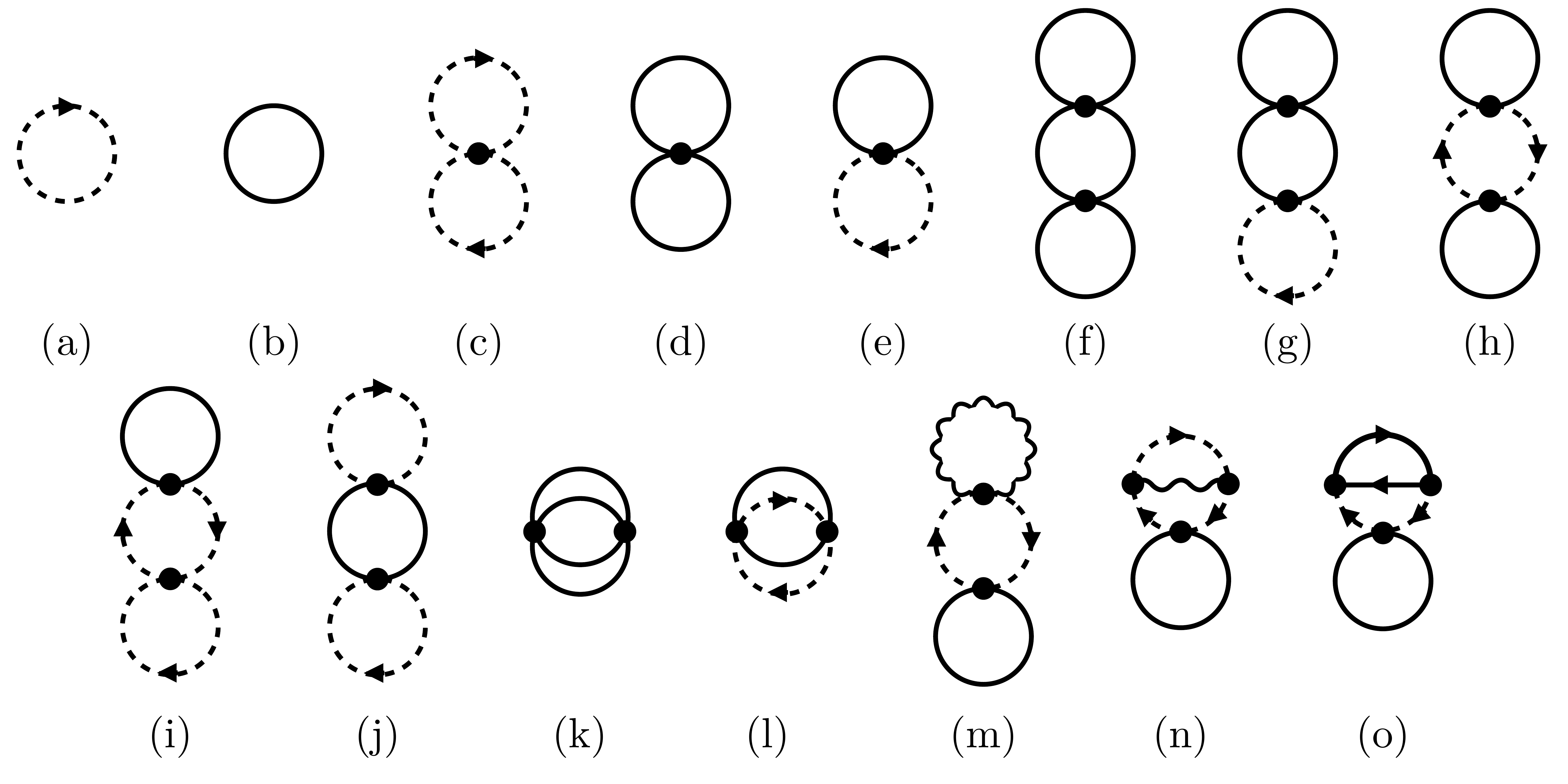} 
\caption{
Diagrams contributing to $ p^{\text{singlet}}_{\text{E}}$. 
Dashed (solid) lines denote the Higgs (singlet), wiggly line gauge fields and the solid arrowed line the top quark.
Vertices in these diagrams are bare vertices, i.e. in renormalised perturbation theory there are counterterm contributions in (a)-(e).
} 
\label{fig:singlet-hard}
\end{figure}

The computation of the singlet contributions depicted in Fig.~\ref{fig:singlet-hard}
is relatively straightforward
-- even at three-loop level --
compared to the computation of the SM contributions: gauge, ghost and fermion topology diagrams do not appear at all at two-loop, and even at three-loop there are only three of them in total. Furthermore, all these reduce trivially to lower order master integrals, since the momentum in the singlet one-loop bubble separates from the other momenta. 
All other three-loop topologies 
with the singlet are pure scalar diagrams of the singlets and Higgs, and therefore straightforward to compute. 
As in Appendix D of Ref.~\cite{Gynther:2005dj} for the SM contributions, we list the explicit midstage results for all different new diagrams containing singlets. 
Note that here we show also pure Higgs diagrams that include counterterms with singlet couplings. 
\begin{align}
(a) &= - d_F  \delta \mu^2_h   \mathcal{I}_1  \; ,  \label{eq:1loop} \\
(b) &= \frac{1}{2} N \Big( \mathcal{I}_0' - (\mu^2_S + \delta \mu^2_S) \mathcal{I}_1 + \frac{1}{2} \mu^4_S \mathcal{I}_2 \Big) \; , \\
(c) &= -d_F (d_F + 1) \delta \lambda_h \mathcal{I}^2_1 \; , \\
(d) &= - \frac{1}{4} N (N+2) (\lambda_S + \delta \lambda_S) \Big( \mathcal{I}^2_1 - 2 \mu^2_S \mathcal{I}_2 \mathcal{I}_1 \Big) \; , \\
(e) &= - \frac{1}{2} d_F N  (\lambda_m + \delta \lambda_m) \Big( \mathcal{I}^2_1 - (\mu^2_h + \mu^2_S) \mathcal{I}_2 \mathcal{I}_1 \Big) 
+ \frac{1}{2} d_F N \lambda_m \delta Z_\phi  \mathcal{I}^2_1  \; , \\
(f) &= \frac{1}{4} N(N+2)^2 \lambda^2_S \mathcal{I}^2_1 \mathcal{I}_2 \; , \\
(g) &=  \frac{1}{2} d_F N(N+2) \lambda_m \lambda_S \mathcal{I}^2_1 \mathcal{I}_2 \; , \\
(h) &= \frac{1}{8} d_F N^2 \lambda^2_m \mathcal{I}^2_1 \mathcal{I}_2  \; , \\
(i) &= N d_F(d_F+1) \lambda_h \lambda_m \mathcal{I}^2_1 \mathcal{I}_2 \; , \\
(j) &= \frac{1}{4} N d^2_F \lambda^2_m \mathcal{I}^2_1 \mathcal{I}_2 \; , \\
(k) &= \frac{1}{4} N(N+2) \lambda^2_S \mathcal{M}_{0,0}  \; , \\
(l) &= \frac{1}{4} d_F N \lambda^2_m \mathcal{M}_{0,0} \; , \\
(m) &= \frac{1}{8} (d_A g^2 + {g'}^2) d_F \lambda_m N (d+\xi) \mathcal{I}^2_1 \mathcal{I}_2 \; , \\
(n) &= \;  - \frac{1}{8}  (d_A g^2 + {g'}^2) d_F \lambda_m N \xi \mathcal{I}^2_1 \mathcal{I}_2,  , \\
(o) &= 2 \lambda_m N N_c g^2_Y \mathcal{I}_1 \widetilde{\mathcal{I}}_1 \mathcal{I}_2 . \label{eq:3looptop} \; .
\end{align}
The results and definitions for the master integrals $\mathcal{I}_{1,2}$, $\widetilde{\mathcal{I}}_1$ and $\mathcal{M}_{0,0}$ can be found in Appendix D of Ref.~\cite{Gynther:2005dj} and Appendix A of Ref.~\cite{Braaten:1995jr}. 
For diagrams with gauge fields, we have used the Fermi gauge, or general covariant gauge, with the identical choice for the SU(2) and U(1) sector gauge fixing parameters
$\xi_2 = \xi_1 = \xi$. 
It can be seen how the gauge fixing parameter $\xi$ cancels for the sum of diagrams (m)$+$(n).
Note that the SM parts in \cite{Gynther:2005dj} were computed in Feynman gauge, and the authors did not explicitly check the gauge invariance of their final result for the pressure. We note that it would be a valuable crosscheck of the correctness of the computation to confirm the gauge invariance of the $p^{\text{SM}}_E(T)$, and for this the automated computation developed and used in Refs.~\cite{Schicho:2020xaf, Croon:2020cgk,Schicho:2021gca} could be utilised.   

From our computation in Eqs.~\eqref{eq:1loop} - \eqref{eq:3looptop}, we find the following results for the different contributions from the hard modes to Eq.~\eqref{eq:psinglet}
\begin{align}
\alpha^{\text{singlet}}_{E1} &= \frac{\pi^2}{90} N , \label{eq:alphaE1}\\ 
\alpha_{E \lambda_m} &= -\frac{d_F}{288} N , \\ 
\alpha_{E \lambda_S} &= - \frac{1}{576} N (N+2) , \\ 
 \alpha_{EA \lambda_m}  &= \frac{C_F d_F N}{144} \bigg(\frac{3}{\epsilon} + 11 + 3 \gamma + 12 \frac{\zeta'(-1)}{\zeta(-1)} + 15 \ln \Big( \frac{\mu}{4 \pi T} \Big) \bigg) , \\ 
\alpha_{EB \lambda_m}  &=  \frac{d_F N}{144} \frac{1}{4} \bigg(\frac{3}{\epsilon} + 11 + 3 \gamma + 12 \frac{\zeta'(-1)}{\zeta(-1)} + 15 \ln \Big( \frac{\mu}{4 \pi T} \Big) \bigg) , \\ 
\alpha_{EY \lambda_m}  &= \frac{N_c N}{72} \bigg(\ln \Big( \frac{\mu}{4 \pi T} \Big) - \ln(2) + \gamma \bigg)  , \\ 
\alpha_{E \lambda_S \lambda_S}  &= \frac{N(N+2)}{144} \bigg( \frac{31}{10} + 6 \frac{\zeta'(-1)}{\zeta(-1)} - 3 \frac{\zeta'(-3)}{\zeta(-3)} + \frac{1}{2} (N+2) \gamma + \frac{1}{2}(N+8) \ln \Big( \frac{\mu}{4 \pi T} \Big) \bigg)  , \\ 
\alpha_{E \lambda_m \lambda_m}  &= \frac{d_F N}{288} \bigg( \frac{31}{5} + 12 \frac{\zeta'(-1)}{\zeta(-1)} - 6 \frac{\zeta'(-3)}{\zeta(-3)} + \frac{1}{2} (N+2d_F) \gamma + \frac{1}{2}(12 + N + 2 d_F) \ln \Big( \frac{\mu}{4 \pi T} \Big) \bigg)  , \\ 
\alpha_{E \lambda_S \lambda_m}  &= \frac{d_F N (N+2)}{144} \bigg(\ln \Big( \frac{\mu}{4 \pi T} \Big) + \gamma \bigg) , \\ 
\alpha_{E \lambda_h \lambda_m}  &= \frac{d_F(d_F+1) N}{72} \bigg(\ln \Big( \frac{\mu}{4 \pi T} \Big) + \gamma \bigg), \\ 
\alpha_{E \mu^2_S} &= -\frac{1}{24} N, \\ 
\alpha_{E \lambda_m \mu^2_S}  &=  \frac{d_F N}{12} \bigg(\ln \Big( \frac{\mu}{4 \pi T} \Big) + \gamma \bigg)  , \\ 
\alpha_{E \lambda_m \mu^2_h}  &=  \frac{d_F N}{12} \bigg(\ln \Big( \frac{\mu}{4 \pi T} \Big) + \gamma \bigg), \\ 
\alpha_{E \lambda_S \mu^2_S}  &=  \frac{N(N+2)}{12} \bigg(\ln \Big( \frac{\mu}{4 \pi T} \Big) + \gamma \bigg) , \\ 
\alpha_{E \mu^4_S}  &= - \frac{N}{2} \bigg( \ln\Big( \frac{\mu_{S}}{4\pi T} \Big) -\frac{3}{4} + \gamma \bigg)  \label{eq:alphaEmus4} .    
\end{align}
We have used similar notation to that of \cite{Gynther:2005dj}, and $L_b$ was defined in Eq.~\eqref{eq:lblf}.
In analogy to \cite{Gynther:2005dj}, the normalisation $p(T = 0) = 0$ in the symmetric phase is accounted for by $\alpha_{E \mu^4_S}$. In practice, we have 
subracted the zero temperature Coleman-Weinberg part from the hard mode contribution 
\begin{align}
\alpha_{E \mu^4_S} = \frac{1}{(4\pi)^2} \frac{1}{4} N \mu^4_S L_b  -\frac{1}{(4\pi)^2} \frac{1}{4} N \mu^4_S \bigg(\ln \Big( \frac{\mu^2_S}{\mu^2} \Big) - \frac{3}{2} \bigg).
\end{align}
Here, the first term is the contribution from the hard modes and the second term is the $T=0$ contribution, c.f. e.g. Eqs.~(A.22) and (2.6) in \cite{Gould:2021oba}. 
Note in particular that the dependence on the RG-scale vanishes in $\alpha_{E \mu^4_S}$.
The divergence related to $\mu^4_S$ is removed by imposing a vacuum counterterm, that reads, together with the similar contribution of the Higgs
\begin{align}
\delta V = -\frac{1}{(4\pi)^2} \frac{1}{4\epsilon}  (N \mu^4_S + 2 \mu^4_h ).
\end{align}
This completes our computation of $p^{\text{singlet}}_{\text{E}}$. 

\subsection{The effective potential, or pressure, in the broken phase}
\label{sec:veff}

The pressure in the broken phases, with non-zero vacuum expectation values for the Higgs or singlet, can be decomposed as 
\begin{align}
p_{}(T) = p_{\text{sym}}(T) - V_{\text{eff}}(v_{\text{min}},s_{\text{min}}), 
\end{align}
where the effective potential is evaluated at its minima $(v_{\text{min}}, s_{\text{min}})$. 
Note that according to our EFT construction, the effective potential is computed in the final ultrasoft scale EFT. 
The parameters, background fields and the ultrasoft RG scale ($\bar{\mu}_3$) of this EFT are denoted by bars.

For the xSM the effective potential has been computed in \cite{Niemi:2021qvp}, in Landau gauge, and here we generalise this computation for $N$ singlets used in this work at hand. We compute the effective potential assuming only one singlet component gets a vev at high temperature, i.e.
$S \rightarrow (S_1+\bar{x} , ..., S_N )^{\text{T}}$. $S_1$ mixes with the Higgs field, and they constitute two mass eigenstates $h_1$ and $h_2$ as in the xSM case, where $N=1$. 
The mixing angle is defined as
\begin{align}
\text{tan} 2 \bar{\theta} \equiv \frac{4 \bar{v} \bar{x} \bar{\lambda}_{m,3}}{2( \bar{\mu}^2_{S,3} - \bar{\mu}^2_{h,3}) + (\bar{\lambda}_{m,3} - 6 \bar{\lambda}_{h,3}) \bar{v}^2 - (\bar{\lambda}_{m,3} - 6 \bar{\lambda}_{h,3}) \bar{x}^2}.
\end{align}
The other $N-1$ inert singlets do not couple to gauge fields, and they merely have interaction with $h_1$, $h_2$, Goldstones and among themselves. All inert singlets have mass (squared) eigenvalue
\begin{align}
\bar{m}^2_S &\equiv \bar{m}^2_{S_i} = \bar{\mu}^2_{S,3} + \bar{\lambda}_{S,3} \bar x^2  + \frac{1}{2} \bar{\lambda}_{m,3} \bar{v}^2, 
\end{align}    
for $2 < i \leq N$. 

In total, we compose the effective potential as 
\begin{align}
V^{\rmii{3d}}_{\text{eff}}(\bar{v},\bar{x}) = V_0 + V_1 + V_2, 
\end{align}
where the indices denote loop-order.
The tree-level potential
\begin{align}
V_0 = V^{\text{xSM}}_0,
\end{align}
is the same as in the xSM.
At one-loop, the inert singlets add a contribution to the 
effective potential
\begin{align}
V_1 = V^{\text{xSM}}_1 + (N-1) J_3(\bar{m}_S),
\end{align}
where
\begin{align}
J_{3}(m_3) \equiv \frac{1}{2} \int_p \ln (p^2 + m^2_3) = -\frac{1}{12 \pi} (m^2_3)^{\frac{3}{2}} + \mathcal{O}(\epsilon). 
\end{align} 
At two-loop, the inert singlets add
\begin{align}
V_2 = V^{\text{xSM}}_2 - \bigg( (\text{SSS}) + (\text{SS}) \bigg),
\end{align} 
where 
\begin{align}
(\text{SSS}) &= (N-1) \bigg( 
  \frac{1}{4} C^2_{s_i s_i h_1} \mathcal{D}_{SSS}(\bar{m}_{S},\bar{m}_{S},\bar{m}_{h,1})
+ \frac{1}{4} C^2_{s_i s_i h_2} \mathcal{D}_{SSS}(\bar{m}_{S},\bar{m}_{S},\bar{m}_{h,2})
\bigg), \\
(\text{SS}) &= 
  (N-1) \bigg( \frac{1}{8} C_{s_i s_i s_i s_i} \Big( I^3_1(\bar{m}_S) \Big)^2
+ \frac{1}{2} C_{s_i s_i G^+ G^-} I^3_1(\bar{m}_S) I^3_1(\bar{m}_G)  \nonumber \\
&+ \frac{1}{4} C_{s_i s_i G G} I^3_1(\bar{m}_S) I^3_1(\bar{m}_G)
+ \frac{1}{4} C_{s_i s_i h_1 h_1} I^3_1(\bar{m}_S) I^3_1(\bar{m}_{h,1}) \nonumber \\
& + \frac{1}{4} C_{s_i s_i h_2 h_2} I^3_1(\bar{m}_S) I^3_1(\bar{m}_{h,2}) \bigg)
+ \frac{1}{2}(N-2)(N-1) \bigg( \frac{1}{4} C_{s_i s_i s_j s_j} \Big( I^3_1(\bar{m}_S) \Big)^2  \bigg)
,
\end{align} 
where the master integrals can be found in the supplementary material of \cite{Niemi:2020hto} 
and the vertex coefficients read
\begin{align}
C_{s_i s_i h_1} &= - \bar{v} \bar{\lambda}_{m,3} \ct + 2 \bar{\lambda}_{S,3} \bar{x} \st , \\
C_{s_i s_i h_2} &= - \bar{v} \bar{\lambda}_{m,3} \st - 2 \bar{\lambda}_{S,3} \bar{x} \ct , \\
C_{s_i s_i s_i s_i} &= -6 \bar{\lambda}_{S,3} , \\
C_{s_i s_i G^+ G^-} &= C_{s_i s_i G G} = -\bar{\lambda}_{m,3} , \\
C_{h_1 h_1 s_i s_i} &= -\bar{\lambda}_{m,3} \ct^2 - 2 \bar{\lambda}_{S,3} \st^2,  \\ 
C_{h_2 h_2 s_i s_i} &= - \bar{\lambda}_{m,3} \st^2 - 2 \bar{\lambda}_{S,3} \ct^2,  \\ 
C_{s_i s_i s_j s_j} &= -2 \bar{\lambda}_{S,3}, \quad (i\neq j),
\end{align} 
where we denote $\ct \equiv \cos \bar{\theta}$ and $\st \equiv \sin \bar{\theta}$. 
The 3d EFT counterterms read
\begin{align}
\delta \bar{\mu}^2_{h,3} &=  \delta \bar{\mu}^2_{h,3,\text{SM}} + \frac{1}{(4\pi)^2} \frac{1}{8\epsilon} N \bar{\lambda}^2_{m,3} , \\
\delta \bar{\mu}^2_{S,3} &=  \frac{1}{(4\pi)^2} \frac{1}{4\epsilon} \Big(  -(3\bar{g}^2_3 + \bar{g}'^2_3)\bar{\lambda}_{m,3}  + 2 \bar{\lambda}^2_{m,3} + 2(N+2)\bar{\lambda}^2_{S,3} \Big) , \\
\delta V_3 &= -\frac{1}{(4\pi)^2} \frac{1}{4\epsilon} (3\bar{g}^2_3 + \bar{g}'^2_3 ) \bar{\mu}^2_{h,3} . 
\end{align}
Note that the vacuum counterterm $\delta V_3$ does not include contributions from the singlet, but since it was not listed in \cite{Niemi:2021qvp}, we add it here. 
The magnetic, or ultrasoft, contribution to the symmetric phase pressure is encoded in the effective potential as $p_{\text{M2}}(T) = T V^{\rmii{3d}}_{\text{eff}}(0,0)$.

\subsection{Order parameters in perturbation theory and gauge invariance}
\label{sec:order-parameter}

\begin{figure}[t]
\centering
\includegraphics[width=0.7\textwidth]{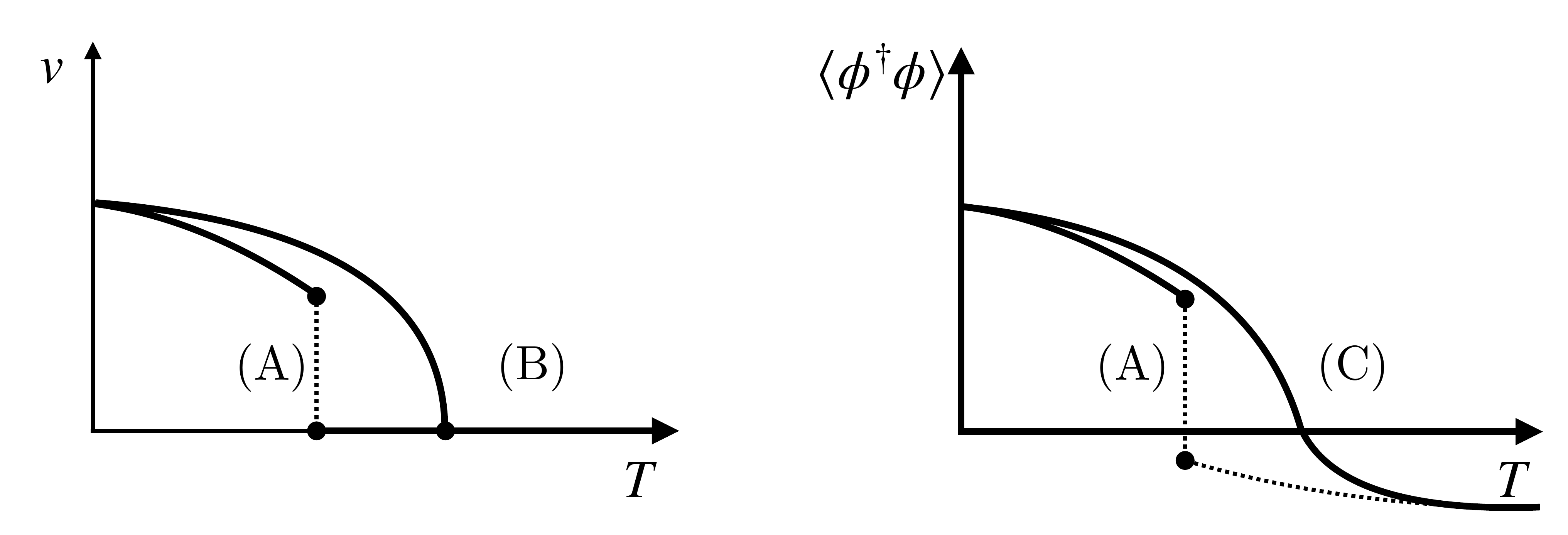}  
\caption{
Schematic illustration of order parameters in perturbation theory, as a function of temperature. 
Left: minimum of the effective potential $(v)$. At high $T$, $v$ is zero, but non-zero at low $T$. If there is a discontinuity, the transition between the two phases is of first order (A). If $v$ is continuous, but there is a discontinuity in the first derivative, the transition is of second order (B).
A smooth crossover (C), where all order derivatives are continuous, is not possible. Furthermore, $v$ itself is not a gauge invariant quantity.
Right: the scalar condensate $\langle \phi^\dagger \phi \rangle$ has a non-vanishing value even in the high $T$ phase, and hence a smooth interpolation between the two phases is possible, leading to a crossover between the phases. Condensates can be computed in a gauge invariant manner, even in perturbation theory. 
} 
\label{fig:order}
\end{figure}

Determining the character of a phase transition -- whether it is of first or second order, or crossover -- is a non-trivial endeavour. In a crossover type transition the system transitions smoothly from one phase to another, and there is no discontinuity in the order parameter at any order in its derivatives. In conventional
analyses in which the minima of the potential are treated as order parameters, such a transition 
can not be recovered:
in the high temperature symmetric phase such minima vanish,
but they 
are non-zero and vary smoothly at lower temperatures,
so there cannot be a smooth interpolation between the phases.
Furthermore, these minima are notoriously gauge dependent,%
\footnote{
Only the value of the effective potential at the minima is gauge invariant, and corresponds to the pressure. 
}
and hence cannot serve as realistic order-parameters.  
More realistic order parameters can be defined in terms of condensates, such as $\langle \phi^\dagger \phi \rangle$ \cite{Farakos:1994xh}. They have a non-zero value even in the symmetric phase
and 

can be computed in a gauge-invariant manner in perturbation theory as 
\begin{align}
\langle \phi^\dagger \phi \rangle = \frac{d V^{\rmii{3d}}_{\text{eff}}}{d\mu^2_3},
\end{align}
where the effective potential is evaluated in an expansion around its leading order minimum. 
Even better, the condensates can be determined in a lattice Monte Carlo simulation, which does not require gauge fixing.
We point out, however, that the computation of these condensates is dependent on their UV renormalisation, and hence their direct numerical values do not have a physical meaning.%
\footnote{
Physical quantities such as the latent heat released in the transition, can be related to these condensates though, see e.g. \cite{Croon:2020cgk}.
}
These order parameter-like quantities are schematically illustrated in Fig.~\ref{fig:order}.

In this article at hand,
we are interested in the sound speed for a strong phase transition, and we have focused solely on the second transition of a two-stepper. In such a case, the barrier separating the phases exists already at tree-level, and the phase transition can hence safely be assumed to be of first order, even without a dedicated non-perturbative study. Regardless, the computation in perturbation theory \textit{should} be arranged in a gauge-invariant manner, in order to study the physical thermodynamic properties of the plasma. Despite this, however, in this article at hand we have resorted to a naive analysis in terms of a direct minimisation of the real part of the potential, %
and treat the location of the minima as a first estimate of the order parameters,
based on the arguments presented in \cite{Niemi:2021qvp}.
With this decision, none of our results are properly gauge invariant.
Furthermore, in \cite{Schicho:2022wty} 
it has been demonstrated that the 
use of either minima \textit{in Landau gauge} or condensates does
not make a qualitative difference.
Hence, we expect our results in Landau gauge to match the gauge invariant analysis, qualitatively.
We argue, that since our main interest is to get a handle on higher order corrections to the speed of sound and estimate their effect on the GW power spectrum, the limitation of a gauge dependent analysis does not compromise the conclusion of this work.
For future work, that would also include a determination of the bubble nucleation rate and the nucleation temperature, and would hence provide a more complete analysis of the predicted GW signature, we envision an upgrade of our current computation of the pressure,%
\footnote{
In our computation, the dimensional reduction step is gauge invariant \cite{Schicho:2021gca}, as well as the symmetric phase pressure \cite{Gynther:2005av, Gynther:2005dj}, but the use of the effective potential is not.
}
in terms of a gauge invariant analysis in perturbation theory, c.f. recent \cite{Laine:2015kra,Gould:2021oba,Croon:2020cgk,Schicho:2022wty} but also \cite{Patel:2011th, Ekstedt:2020abj}. In addition, in such an analysis, a consistent perturbative expansion around the leading order minima, imaginary parts do not show up.
The effective potential itself has a spurious imaginary part, that we have simply disregarded in our direct minimisation of the potential.

\subsection{Relation to lower order computations}
\label{sec:4d-veff}

For readers not familiar with the 3d EFT approach, we describe here the commonly adopted one-loop computation, in order to provide a comparison.
For the remainder of this section, we set $N=1$, i.e. we discuss the xSM. In a one-loop computation, the EFT technology is not as apparent as in computations required for higher order corrections. 
Regardless, even the one-loop effective potential based on daisy resummation by Arnold and Espinosa \cite{Arnold:1992rz} utilises the EFT picture for the thermal scale hierarchy, as only the soft modes are resummed, corresponding to screening of the hard scale. However, the underlying EFT picture is not apparent  
in \cite{Arnold:1992rz}
and the computation is performed directly in the 4d parent theory;
a formal discussion in the EFT language is formulated in \cite{Kajantie:1995dw,Braaten:1995cm}.
At leading order, 
the parameters of the EFT have trivial relations to the 4d parameters, i.e. $\lambda_3 \sim T \lambda$ and $v^2_3 \sim v^2/T$, and only the masses receive thermal corrections, at one-loop level. 
Let us start by taking a closer look at the all-order resummation of the \textit{leading} daisy diagrams. 

\begin{figure}[t]
\centering  
\includegraphics[width=0.8\textwidth]{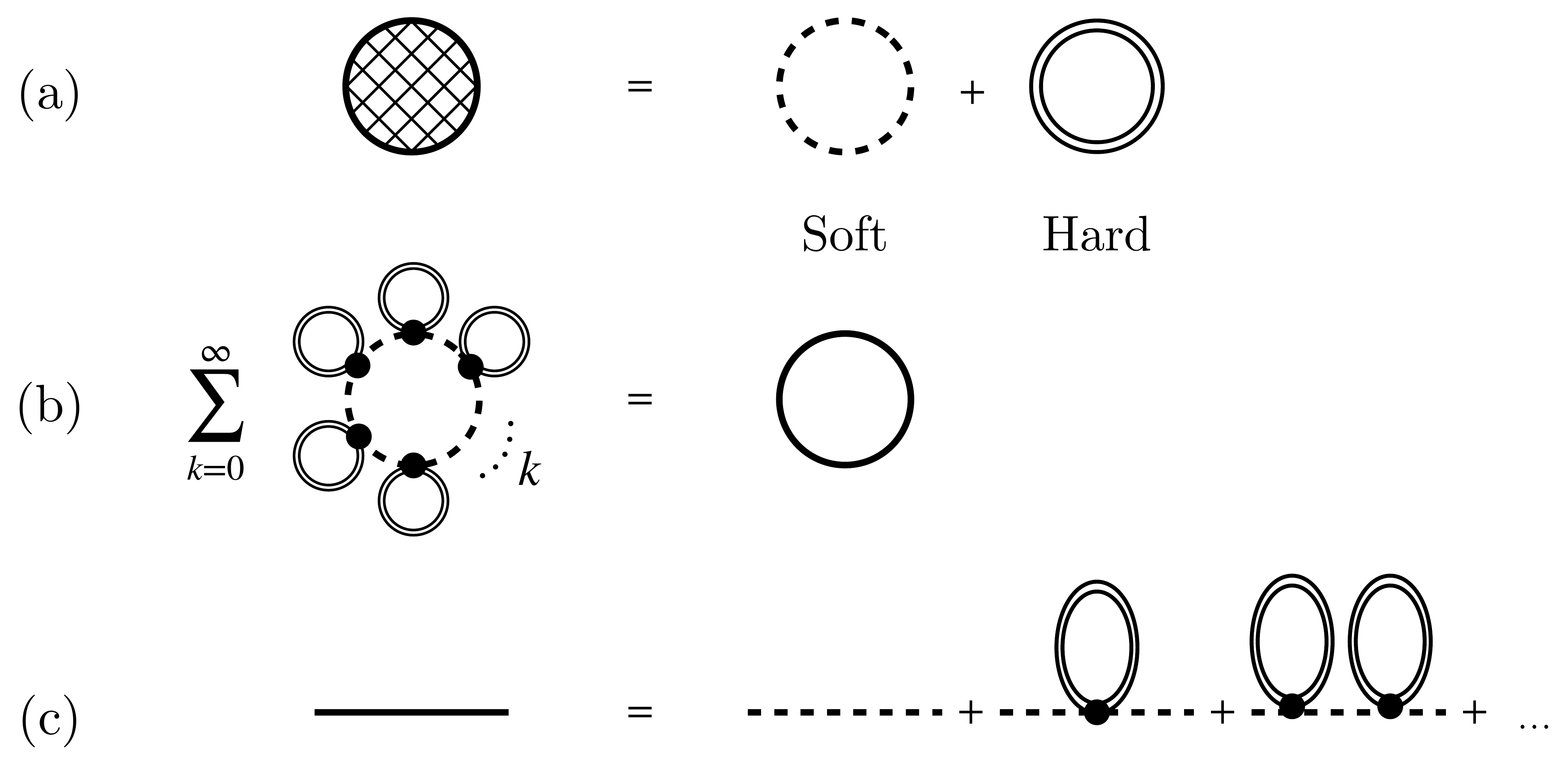}   
\caption{
Schematic illustration of leading daisy resummation. 
(a): division of the one-loop vacuum bubble into soft (dashed line) and hard contributions (double lines).
(b): all-order resummation of the hard mode contributions to the soft term.
(c): resummation of the soft scalar mass parameter. 
} 
\label{fig:daisy}
\end{figure}

\paragraph{Leading order daisy resummation}
In Fig.~\ref{fig:daisy} we illustrate schematically the premise of daisy resummation. 
Fig.~\ref{fig:daisy} (a) showcases the division of the one-loop vacuum bubble into soft (dashed line) and hard contributions (double lines). While the hard modes are regulated in the IR by non-zero Matsubara frequencies, the soft part is IR-sensitive. 
Physically, this is related to macroscopic collective phenomena, i.e. screening of different scales in the plasma. Technically, a problem appears in loops including the zero-mode, since the propagator has the same form as at $T=0$, while the integration measure has one less power of momentum: this leads to worse behaviour in the IR compared to the $T=0$ case. 
This is problem is demonstrated in Fig.~\ref{fig:daisy} (b): consider a $k$-loop \textit{daisy diagram} where the inner loop has a soft loop-momentum with $k$ propagators with \textit{unresummed} mass $m^2 \sim (g T)^2 $, and $k$ loops  with hard loop momenta $P$.
When each vertex contributes with $g^2$ this diagram has the form \cite{laine-thesis} 
\begin{align}
&\sim \bigg[ T \int_p \frac{1}{(p^2+m^2)^k} \bigg]_{\text{soft}} \bigg[ g^2 \sumint{P}' \frac{1}{P^{2}} \bigg]^k_{\text{hard}} \nonumber \\
&= \bigg[ T \frac{1}{(-1)_{k-1}} \frac{d^{k-1}}{d(m^2)^{k-1}} \int_p \frac{1}{(p^2+m^2)} \bigg]_{\text{soft}} \bigg[ \frac{1}{12^k} g^{2k} T^{2k}  \bigg]_{\text{hard}} \nonumber \\
&= \bigg[ -\frac{T}{4\pi}  \frac{(\frac{1}{2})_{k-1}}{(-1)_{k-1}} (m^2)^{\frac{3}{2}-k} \bigg]_{\text{soft}} \bigg[ \frac{1}{12^k} g^{2k} T^{2k}  \bigg]_{\text{hard}} \nonumber \\
&= \bigg( -\frac{1}{4\pi \times12^k} \frac{(\frac{1}{2})_{k-1}}{(-1)_{k-1}} \bigg) \bigg( m^3 T \Big( \frac{g T}{m} \Big)^{2k} \bigg) \sim g^3 \quad \text{for all} \; k.
\end{align}  
Results for the used one-loop integrals can be found in e.g. \cite{Braaten:1995jr}. 
Here $(x)_n$ denotes the falling factorial, but the overall numerical constant is irrelevant for the observation, that regardless of $k$, any such diagram contributes at $\mathcal{O}(g^3)$, since $m \sim g T$. Even worse, for $k\geq2 $ and $m \rightarrow 0$ all these contributions are IR-divergent.
However, for massive fields  -- i.e. scalars and temporal scalar fields, but not for massless, spatial gauge fields -- there is a salvation 
by means
of a resummation. In Fig.~\ref{fig:daisy} (c) the soft scalar propagator is resummed as a geometric Dyson series. Formally, the scalar 2-point Green's function $G$ can be witten in terms of the tree-level propagator $G_0 = 1/(p^2 + m^2)$ and the \textit{one-loop} thermal correction $\Pi \sim g^2 T^2$ as
\begin{align}
G &= G_0 + G_0 \Pi G_0  +  G_0 (\Pi G_0 \Pi) G_0 + \ldots 
= G_0 \Big( \sum_{i=0}^{\infty} (\Pi G_0)^i \Big) \nonumber \\
&=  G_0 \Big( \frac{1}{1-\Pi G_0} \Big)   = \frac{1}{p^2 + m^2 + \Pi},  
\end{align}     
which leads to a thermally corrected mass for the three-dimensional zero-mode $m^2_3 \equiv m^2 + \Pi$, which is resummed to all-orders, by the one-loop hard correction. When all different daisy diagrams in Fig.~\ref{fig:daisy} (b) are summed together, the result for the one-loop soft vacuum bubble reads
\begin{align}
(\text{LO soft}) = -\frac{m^3_{3}}{4\pi} \sim \mathcal{O}(g^3).
\end{align}
When this contribution is combined with the tree-level terms and thermally corrected mass at $\mathcal{O}(g^2)$, the effective potential is consistent at $\mathcal{O}(g^3)$. 

\begin{figure}[t]
\centering  
\includegraphics[width=0.45\textwidth]{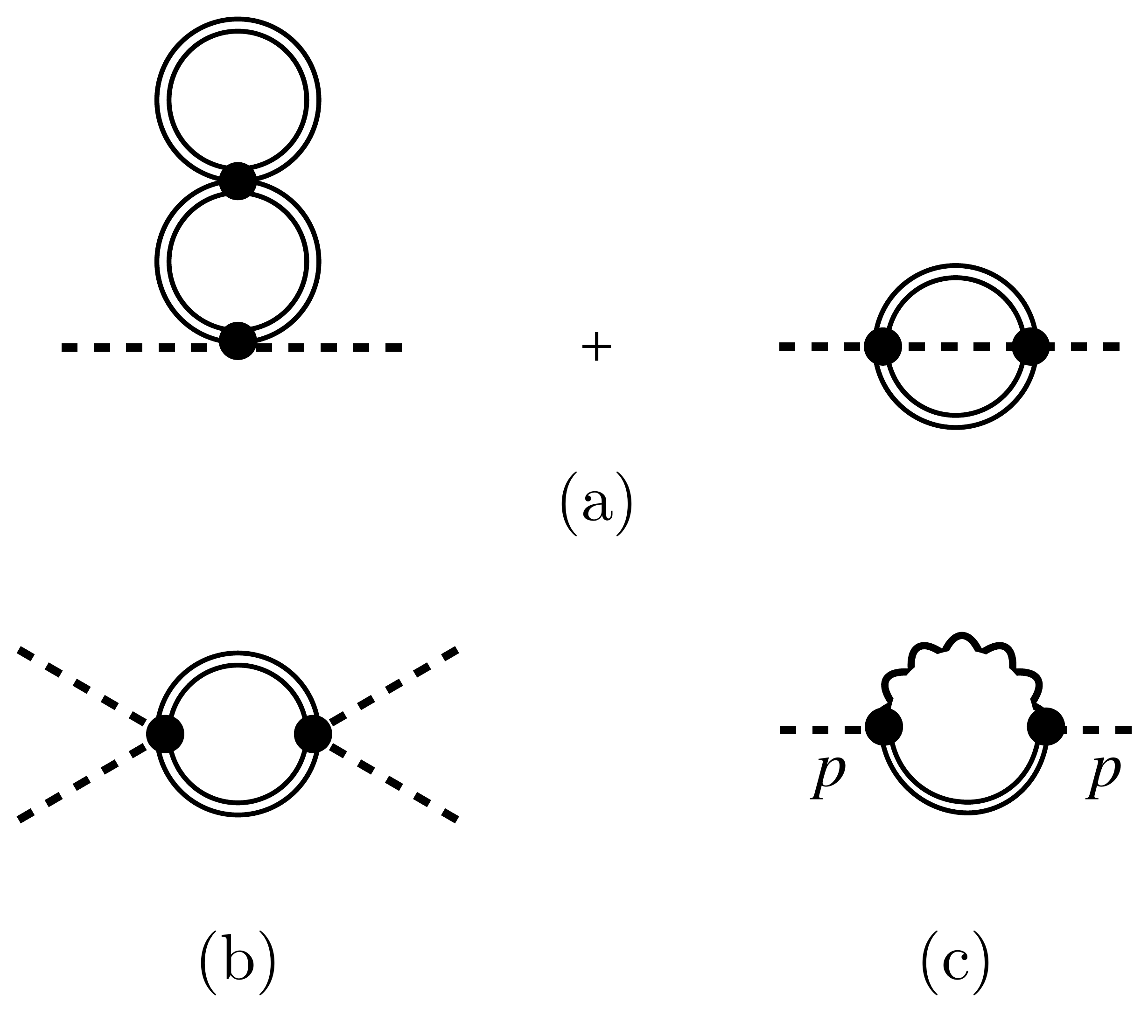}   
\caption{
Illustration of higher order resummations in NLO dimensional reduction, going beyond leading one-loop daisy resummation of the mass. 
(a): two-loop resummation of the mass parameter.
(b): one-loop resummation of the couplings.
(c): one-loop resummation of the fields, from the part 
dependent on the external momentum.
Here a dashed line represents a soft field, and a double line and wiggly line hard scalar and gauge field modes, respectively.
All NLO effects in dimensional reduction contibute at $\mathcal{O}(g^4)$. 
} 
\label{fig:DR}
\end{figure}

In Fig.~\ref{fig:daisy} (a) in addition to the $\mathcal{O}(g^3)$ soft contribution, there are
one-loop hard mode contributions -- that include the $T=0$ Coleman-Weinberg potential (see end of this section) -- which appear at order $\mathcal{O}(g^4)$. Alas, for a complete computation at this order, also higher order resummations 
are required. Such higher order resummations are those depicted in Fig.~\ref{fig:DR}: (a) two-loop corrections to the mass, (b) one-loop corrections to couplings and (c) one-loop resummation of the fields, from momentum dependent contributions to the 2-point function. The inclusion of all these contributions is systematic in the EFT language, and they are resummed to all-orders in dimensional reduction to 3d EFT, in analogy to daisy resummation that resums one-loop thermal mass to all orders.    

\paragraph{Comparison of 3d and 4d effective potentials}
Schematically, the relation between the lower-order 4d effective potential and the higher-order 3d effective potential is%
\footnote{Here the labels 4d and 3d merely denote the approach used to compute the free energy, that describes physical properties of the plasma.}
\begin{align}
V^{\text{4d}}_{\text{eff}}(v,s) \simeq T \; V^{\text{3d}}_{\text{eff}}(v_3,s_3).
\end{align}
This relation is only an approximate equality, since $V^{\text{3d}}_{\text{eff}}(v_3,s_3)$ contains higher-order contributions.
Technically, the one-loop $V^{\text{4d}}_{\text{eff}}$ with daisy resummation is correct at $\mathcal{O}(g^3)$, but incomplete at $\mathcal{O}(g^4)$. 
By comparing the above expressions, 
we have for the soft terms
\begin{align}
V^{\text{4d}}_{\text{soft}} = T \; V^{\text{3d}}_{\text{1-loop}},
\end{align}
which holds exactly if only the LO matching relations of Eqs.~\eqref{eq:LO-DR-1} and \eqref{eq:LO-DR-2} are used, at $\mathcal{O}(g^2)$.
In practice one uses $\mathcal{O}(g^4)$ accurate relations in the 3d EFT, providing higher accuracy.
We emphasise, that this is exactly the term that corresponds to leading daisy resummation. 
For the hard terms we have 
\begin{align}
\label{eq:compare-hard}
V^{\text{4d}}_{\text{tree}}+ V^{\text{4d}}_{\text{hard}} \simeq T \; V^{\text{3d}}_{\text{tree}}.
\end{align}
Again, equality holds exactly if the expressions are computed to the same order in the coupling expansion. In the 3d picture, hard contributions are encoded in the parameters of the EFT.
Here we emphasise, that in the 4d approach, the leading contributions that depend on the external momenta -- corresponding to a screening of the soft field by the heat bath -- are not included.
In the 3d computation these terms are included, and necessary for a consistent computation at $\mathcal{O}(g^4)$ and RG-improvement, see Ref.~\cite{Arnold:1992rz,Farakos:1994kx,Gould:2021oba}.
As discussed in these references, at $\mathcal{O}(g^4)$ also two-loop contributions to the thermal masses 
are required, as well as a two-loop effective potential for the soft or 3d terms.

For concreteness, below we construct again the one-loop effective potential using the background field method.
We need the mass squared eigenvalues for fluctuating quantum fields
\begin{align}
m^2_\chi &= \mu^2_h + \lambda_h v^2 + \frac{1}{2} \lambda_m s^2, \\  
m^2_{\pm} &= \frac{1}{4} \bigg\{ 2 \mu^2_h + 2 \mu^2_{S} + (6\lambda_h + \lambda_m )v^2 + (\lambda_m + 6\lambda_{S} ) s^2 \pm \sqrt{ A + B } \bigg\}, \\
A &\equiv (-6\lambda_h+\lambda_m)^2 v^4 + \Big( 2 \mu^2_h - 2 \mu^2_{S} + (\lambda_m - 6\lambda_{S}) s^2  \Big)^2, \\
B &\equiv 2 v^2 \bigg( 2 (6\lambda_h-\lambda_m)(\mu^2_h -\mu^2_{S}) + \Big( 6 \lambda_h ( \lambda_m - 6 \lambda_{S}  ) + \lambda_m ( 7 \lambda_m + 6 \lambda_{S} ) \Big) s^2 \bigg),
\end{align}     
where $m_{\pm}$ are the two neutral scalar mass eigenvalues, and the Goldstone mass eigenvalue $m^2_\chi$ is triply-degenerate. 
For the gauge field and top quark pieces, we need  the
mass eigenvalues
\begin{align}
m^2_W = \frac{1}{4} g^2 v^2, \quad \quad
m^2_Z = \frac{1}{4} (g^2 + {g'}^2) v^2, \quad \quad
m^2_t = \frac{1}{2} g^2_Y v^2.
\end{align}
The mass squared eigenvalues related to the temporal scalar fields read \cite{Farakos:1994kx,Kajantie:1995dw} 
\begin{align}
M^2_3 &= m^2_D + h_3 v^2_{3} , \\  
M^2_{\pm,3} &= \frac{1}{2} \bigg\{ m^2_D + m_D'^2 + (h_3 + h_3') v^2_{3} + \pm \sqrt{ X + Y } \bigg\}, \\
X &\equiv (m_D^2 - m_D'^2)^2 - 2 (h_3 - h_3')(m_D'^2 - m_D^2) v^2_3 , \\
Y &\equiv \Big( (h_3 - h_3')^2 + h''^2_{3}  \Big) v^4_3,
\end{align}   
where $M^2_3$ 
is doubly-degenerate. 
Here all quantities are those of the 3d EFT, since temporal scalars live at the soft scale.%
\footnote{
Temporal scalars $A^a_0$ and $B_0$ are mixing through a $h_3''$ term, but in practise one can linearise this mixing.
}
Coefficients $h_3, h_3', h_3''$ are couplings between Higgs and temporal scalars, see \cite{Kajantie:1995dw, Schicho:2021gca}.

Above, the mass eigenvalues are input for the one-loop master integral 
\begin{align}
V^{\text{4d}}_{\text{1-loop}} &\sim \frac{1}{2} \sumint{P} \ln (P^2 + m^2) = J_{\text{soft}}(m) + J_{\text{hard}}(m).
\end{align}
We divide it into the soft zero-mode contribution
\begin{align}
J_{\text{soft}}(m_3) &= T J_{3}(m_3), 
\end{align}
with all quantities resummed, i.e. those of the 3d EFT, and denoted herein with subscript 3, 
and contributions from the hard non-zero modes
\begin{align}
J_{\text{hard}}(m) &\equiv \frac{1}{2} \sumint{P}' \ln (P^2 + m^2) = \frac{1}{2} \sumint{P}' \ln P^2 + \frac{1}{2} \sum_{n=1}^{\infty} \frac{(-1)^{n-1}}{n} (m^2)^n \sumint{P}' \frac{1}{(P^2)^n},
\end{align}
where we have used the high-$T$ expansion in $m^2/T^2$ for the last line. 
Concretely 
\begin{align}
J^b_{\text{hard}} &= -\frac{\pi^2}{90} T^4 + \frac{1}{24} T^2 m^2 - \frac{m^4}{4(4\pi)^2} \Big( \frac{1}{\epsilon} + L_b \Big) + \mathcal{O}\Big( \frac{m^6}{T^2} \Big), \\ 
J^f_{\text{hard}} &= -\frac{7}{8} \frac{\pi^2}{90} T^4 - \frac{1}{48} T^2 m^2 - \frac{m^4}{4(4\pi)^2} \Big( \frac{1}{\epsilon} + L_f \Big) + \mathcal{O}\Big( \frac{m^6}{T^2} \Big) \; ,
\end{align}
where we give the result for both bosonic and fermionic fields.
Typically only the first three terms are kept. Higher order terms would result in higher dimensional, marginal, operators in the 3d EFT \cite{Kajantie:1995dw}.

In total, the one-loop effective potential reads (in Landau gauge) 
\begin{align}
V^{\text{4d}}_{\text{eff}} = V^{\text{4d}}_{\text{tree}} + V^{\text{4d}}_{\text{CT}} + V^{\text{4d}}_{\text{1-loop}},  
\end{align}
where
\begin{align}
V^{\text{4d}}_{\text{tree}} &= \frac{1}{2} \mu^2_{h} v^2 + \frac{1}{4} \lambda_{h} v^4 + \frac{1}{2} \mu^2_{S} s^2 + \frac{1}{4} \lambda_{S} s^4 + \frac{1}{4} \lambda_{m} v^2 s^2, \\
V^{\text{4d}}_{\text{CT}} &= \frac{1}{2} \delta\mu^2_{h} v^2 + \frac{1}{4}  \delta\lambda_{h} v^4 + \frac{1}{2}  \delta\mu^2_{S} s^2 + \frac{1}{4}  \delta\lambda_{S} s^4 + \frac{1}{4}  \delta\lambda_{m} v^2 s^2,
\end{align}
and
\begin{align}
V^{\text{4d}}_{\text{1-loop}} = V^{\text{4d}}_{\text{soft}} + V^{\text{4d}}_{\text{hard}}, 
\end{align}
with ($d=3-2\epsilon$)
\begin{align}
V^{\text{4d}}_{\text{hard}} &= 3 J^b_{\text{hard}}(m_\chi) + J^b_{\text{hard}}(m_+) + J^b_{\text{hard}}(m_-) \nonumber \\
& + d \Big( 2 J^b_{\text{hard}}(m_W) + J^b_{\text{hard}}(m_Z) \Big) - 4 N_c J^f_{\text{hard}}(m_t), \\
V^{\text{4d}}_{\text{soft}} &= 3 J_{\text{soft}}(m_{\chi,3}) + J_{\text{soft}}(m_{+,3}) + J_{\text{soft}}(m_{-,3}) \nonumber \\
& + (d-1) \Big( J_{\text{soft}}(m_{W,3}) + J_{\text{soft}}(m_{Z,3}) \Big) \nonumber \\
& + 2 J_{\text{soft}}(M_3) + J_{\text{soft}}(M_{+,3}) + J_{\text{soft}}(M_{-,3}). 
\end{align}
The counterterms cancel the $1/\epsilon$ poles 
in $J_{\text{hard}}$. 
In the soft pieces, the conventional daisy-resummed approach corresponds to utilising the dimensional reduction matching relations at leading order, i.e. one-loop for the masses and tree-level for the couplings
\begin{align}
\label{eq:LO-DR-1}
\mu^2_{h,3} &= \mu^2_h + \Pi_\phi, \quad \quad \mu^2_{S,3} = \mu^2_S + \Pi_S, \\
\label{eq:LO-DR-2}
\lambda_{h,3} &= T \lambda_h, \quad \quad \lambda_{S,3} = T \lambda_S \quad \quad \lambda_{m,3} = T \lambda_m.
\end{align}
The background fields are related as $v_3 = v/\sqrt{T}$ and $s_3 = s/\sqrt{T}$.
The one-loop thermal mass contributions read
\begin{align}
\Pi_\phi &= \frac{T^2}{12} \Big( \frac{3}{4}(3g^2 + {g'}^2) +3 g^2_Y + 6 \lambda_h + \frac{1}{2} \lambda_m \Big) , \\
\Pi_S &= \frac{T^2}{12} \Big( 2 \lambda_m + 3 \lambda_S \Big) , \\
m^2_D &= g^2 T^2 \Big(\frac{5}{6} + \frac{1}{3} N_f \Big), \\
m_D'^2 &= {g'}^2 T^2 \Big(\frac{1}{6} + \frac{5}{3} N_f \Big).
\end{align}
This completes the construction of the one-loop effective potential. Note that the zero temperature Coleman-Weinberg potential is implicitly included above, as 
\begin{align}
\label{eq:veff4d}
V^{\text{4d}}_{\text{1-loop}} &\sim \frac{1}{2} \sumint{P} \ln (P^2 + m^2) = J_{\text{soft}}(m) + J_{\text{hard}}(m)
= J_{\text{CW}}(m) + J_{T}(m), 
\end{align}
where
\begin{align}
J_{\text{CW}}(m) &= \frac{1}{2} \Big( \frac{\mu^2 e^\gamma}{4\pi} \Big)^\epsilon \int \frac{d^{D}p}{(2\pi)^D} \ln(p^2+m^2)  = -\frac{1}{2}
  \Big( \frac{\mu^2 e^\gamma}{4\pi} \Big)^\epsilon
  \frac{ \Big(m^2 \Big)^\frac{D}{2}}{(4\pi)^{\frac{D}{2}}} \frac{\Gamma(-\frac{D}{2})}{\Gamma(1)} \; , \\ 
J_{T}(m) &= - \int_{p} T \ln \Big( 1 \pm n_{\rmii{B/F}}(E_p,T)  \Big).  
\end{align}
Here $J_{\text{CW}}$ is the result at $T=0$ in $D=d+1=4-2\epsilon$ dimensions and all temperature dependence is captured in the thermal function $J_{T}$%
\footnote{This separation can be obtained by a contour trick \cite{Kapusta:2006pm,Laine:2016hma}, where the sum over Matsubara frequencies is turned into a contour integration in the complex plane.} 
in which $n_{\rmii{B/F}}$ is the Bose or Fermi distribution.
Note, however, that in these expressions the daisy resummation is not yet implemented.
To re-express Eq.~\eqref{eq:veff4d} into the more popular form with daisy resummation, we write
\begin{align}
J_{\text{CW}}(m) + J_{T}(m) 
&= J_{\text{CW}}(m) + \underbrace{\Big(J'_{T}(m) + J_{\text{soft}}(m)  \Big)}_{J_T} \nonumber \\
&\rightarrow   J_{\text{CW}}(m) + J'_{T}(m) + J_{\text{soft}}(m_3) \nonumber \\ 
&= J_{\text{CW}}(m) + J_{T}(m) + \underbrace{\Big( J_{\text{soft}}(m_3) - J_{\text{soft}}(m)   \Big)}_{\equiv J_{\text{daisy}}} \nonumber \\
&= J_{\text{CW}}(m) + J_{T}(m) +  J_{\text{daisy}}(m_3,m),  
\end{align}
a form often encountered in the literature \cite{Quiros:1999jp}.
Here the prime denotes the sole hard contributions, and the resummation is implemented by $J_{\text{soft}}(m) \rightarrow J_{\text{soft}}(m_3)$, i.e. resumming the soft mode. 
The arguably counterintuitive form $J_{\text{daisy}}(m_3,m) \equiv J_{\text{soft}}(m_3) - J_{\text{soft}}(m)$ simply results from the fact that the unresummed zero mode needs to be subtracted, as it is still present in $J_{T}(m)$. Despite this counterintuitive formulation, by nature this resummation is still equal to LO dimensional reduction and use of the 3d EFT.    
The unresummed thermal function $J_T(m)$ can be evaluated numerically without using the high-$T$ expansion, while the resummation of the zero mode is encoded in $J_{\text{daisy}}(m_3,m)$. 
However, the zero mode is resummed due to screening by hard non-zero modes, and still assumes the high temperature scale hierarchy.
Therefore, in this formulation it is \textit{not} straightforward to move to a low-temperature limit even if the thermal function is handled numerically, without high temperature expansion.  

For completeness, we list high-$T$ expansions of bosonic and fermionic thermal functions in a form often encountered in the literature
\begin{align}
J_{T,b}(m) = T^4 \bigg( -\frac{\pi^2}{90}  + \frac{1}{24} \frac{m^2}{T^2} - \frac{1}{4(4\pi)^2} \frac{m^4}{T^4} \ln \Big(\frac{m^2}{a_b T^2}\Big) + \mathcal{O}\Big( \frac{m^6}{T^6} \Big) \bigg), \\ 
J_{T,f}(m) = T^4 \bigg( -\frac{7}{8} \frac{\pi^2}{90}  -\frac{1}{48} \frac{m^2}{T^2} - \frac{1}{4(4\pi)^2} \frac{m^4}{T^4} \ln \Big(\frac{m^2}{a_f T^2}\Big) + \mathcal{O}\Big( \frac{m^6}{T^6} \Big) \bigg),
\end{align}
where $a_b = (4\pi)^2 \text{Exp}(\frac{3}{2}-2\gamma)$ and $a_f = a_b/16$. 
In the high-$T$ expansion it is straightforward to check that indeed $J_{\text{CW}}+J_{T}' = J_{\text{hard}}$,
i.e. the zero-$T$ Coleman-Weinberg potential is included in $J_{\text{hard}}$. 

Finally, we comment that above we did not yet include $\# T^4$ contributions from gluons and light fermions in the effective potential, but this can be done as in Sec.~\ref{sec:p-sym} for the leading order pressure.

{\small
%
\bibliographystyle{bib/JHEP.bst}
\bibliography{bib/references.bib}

}
\end{document}